\documentclass[11pt,a4paper]{article}
\usepackage{jheppub}
\usepackage{bm}
\usepackage{hyperref}
\newcommand{\diag}{{\rm diag}}

\newcommand{\ssA}{{\scriptscriptstyle{ A}}}
\newcommand{\ssV}{{\scriptscriptstyle{ V}}}
\newcommand{\ssD}{{\scriptscriptstyle{ D}}}

\newcommand{\ssR}{{\scriptscriptstyle{ R}}}

\newcommand{\ssF}{{\scriptscriptstyle{ F}}}
\newcommand{\ssE}{{\scriptscriptstyle{ E}}}
\newcommand{\ssB}{{\scriptscriptstyle{ B}}}
\newcommand{\ssL}{{\scriptscriptstyle{ L}}}

\newcommand{\ssM}{{\scriptscriptstyle{ M}}}
\newcommand{\ssC}{{\scriptscriptstyle{ C}}}

\newcommand{\ssQ}{{\scriptscriptstyle{ Q}}}

\newcommand{\sschi}{{\scriptscriptstyle{ \chi}}}

\newcommand{\ssa}{{\scriptscriptstyle{a}}}
\newcommand{\ssm}{{\scriptscriptstyle{m}}}

\newcommand{\ssBb}{{\scriptscriptstyle{ B5}}}
\newcommand{\ssbes}{{\scriptscriptstyle{ 5}}}

\newcommand{\mnar}{{\scriptscriptstyle{\mu \nu \alpha \rho }}}

\newcommand{\ssmn}{{\scriptscriptstyle{\mu \nu }}}

\newcommand{\ssbfn}{{\scriptscriptstyle{(n)}}}
\newcommand{\ssbfu}{{\scriptscriptstyle{(u)}}}

\newcommand{\tN}{{\tilde{\nabla}}}

\long\def\symbolfootnote[#1]#2{\begingroup%
	\def\thefootnote{\fnsymbol{footnote}}\footnote[#1]{#2}\endgroup}

\begin{document}
	
\title{Quantum Kinetic Equation for Fluids of Spin-$\mathbf{1/2}$ Fermions}

\author[a]{\"{O}mer F. Dayi}
\author[b]{Eda Kilin\c{c}arslan}

\affiliation[a]{Physics Engineering Department, Faculty of Science a nd
	Letters, Istanbul Technical University,
	TR-34469, Maslak--Istanbul, Turkey}
\affiliation[b]{Department of Physics, Bogazici University, 34342 Bebek, Istanbul, Turkey}

\emailAdd{dayi@itu.edu.tr }

\emailAdd{edakilincarslann@gmail.com}



\abstract{Fluid of spin-1/2 fermions is represented by a complex scalar field and a four-vector field coupled both to the scalar and the Dirac fields. We present the underlying action and show that  the resulting equations of motion are identical to the (hydrodynamic) Euler equations in the presence of Coriolis force. As a consequence of the gauge invariances of this action we established the  quantum kinetic equation which takes account of noninertial properties of the fluid in the presence of electromagnetic  fields. The  equations of the field components  of Wigner function in Clifford algebra basis are employed to construct new semiclassical covariant  kinetic equations of the vector and axial-vector field components for massless as well as   massive fermions. Nonrelativistic limit of the chiral kinetic equation is studied and shown that it generates a novel three-dimensional transport theory which does not depend on spatial variables explicitly and possesses a Coriolis force term. We demonstrated that the three-dimensional chiral transport equations are consistent with the chiral anomaly. For massive fermions the three-dimensional  kinetic transport theory generated by the new covariant kinetic equations is established in small mass limit. It possesses the Coriolis force and the massless limit can be obtained directly.}

	\maketitle
\section{Introduction}
\label{intro}
Wigner function has been originally introduced in deformation quantization of classical mechanics as the substitute of probability density in ordinary quantum mechanics. However, it may take negative values, so that it is considered as the quasiprobability  distribution which provides quantum corrections to classical statistical mechanics. It is built with the wave functions satisfying Schroedinger equation \cite{Wig-rev}. This quantum mechanical formulation can be generalized to field theory by employing  quantum fields to construct  the Wigner function \cite{Groot}. For  spin-$1/2$ fermions the constituent fields  satisfy the Dirac equation. It turned out that the Wigner function method is a powerful tool in constructing relativistic quantum kinetic theories of spin-$1/2$ fermions \cite{Heinz83-1, egv,vge,zh1}.

In heavy-ion collisions quark-gluon plasma is created where the constituent quarks are approximately massless \cite{Gyulassy:2004zy, Shuryak:2004cy}. Their properties can be inspected within the chiral kinetic theory (CKT) which leads to an intuitive understanding of  chiral magnetic effect (CME) \cite{kmw,fkw,kz} and chiral separation effect (CSE) \cite{mz,jkr} for the chiral plasma subjected to the  external electromagnetic fields. When the chiral plasma is considered as a fluid, the chiral transport equations are  also useful to study the chiral vortical effect (CVE) \cite{ss} and local polarization effect (LPE) \cite{lw,bpr,glpww}.

The  main characteristics of  the quarks in quark-gluon plasma are acquired by considering them as massless particles. However, it as an approximation. Therefore, studying the mass corrections is needed. The Wigner function formalism  of massive spin-$1/2$ fermions yields some different covariant kinetic equations  \cite{wsswr,hhy}, in contrary to the massless case.  This is due to the fact that for massive fermions   there are more than one way of eliminating the irrelevant set of field equations derived from the quantum kinetic equation (QKE). 

Although four-dimensional (4D) approach has some  advantages like being manifestly Lorentz invariant, non-relativistic or equal-time formalism  is needed if one would like to solve a transport equation starting from an initial distribution function provided by  field equations \cite{bbgr,zh2,oh}. 
 Integrating 4D transport equation  over the zeroth-component  of momentum is the custom method of constructing the related  three-dimensional (3D)  transport equation \cite{zh1,oh}. 

Wigner function of spin-$1/2$ fermions which are coupled to gauge fields  is constructed to be invariant under gauge transformations which leave the Dirac equation intact.  The gauge invariant Wigner function satisfies the QKE which depends on the  field strength explicitly \cite{egv,vge}. When one deals with relativistic plasma as a fluid, the vorticity effects should also be taken into account.  Although the magnetic and vortical effects are similar,  the QKE does not explicitly depend  on the  rotational properties of fluid in contrary to electromagnetic interactions. Hence,  noninertial effects  like the Coriolis force are absent.  To overcome these difficulties we proposed the modification of QKE by means of  enthalpy current either for the massless or massive fermions \cite{dk,dk-m}. We obtained the relativistic transport equations and studied the 3D theories which they generate. The chiral formulation was successful in generating a consistent 3D CKT which does not depend explicitly on the position vector and also addresses the noninertial effects like the Coriolis force correctly. The modification of QKE also gives satisfactory results for massive   fermions. However, the modified  QKE  has not been following from an action in contrary to the electromagnetic part. Here, we present an underlying Lagrangian which naturally yields  the aforementioned modification of QKE.

3D CKT with the Coriolis force was first presented in \cite{SteYin} by making use of the resemblance between the magnetic field and  angular velocity. Then this formulation was derived in \cite{husad} from the first principles in a rotating coordinate frame. In  \cite{husad} it was also shown that spatial coordinate dependence appearing  in  some CKTs can be removed by an appropriate phase space coordinate transformations. 

{ In \cite{lgmh} CKT in curved spacetime has been  derived from the QKE. There it was shown that  the noninertial effects and the CVE arise when the observer is in  the comoving frame of fluid \cite{lgmh}. In this case  our modification terms vanish identically  as we have already  discussed in detail  in \cite{dk2019}.   However,  in flat spacetime formalism,  to generate the noninertial effects one should consider the modified QKE. There is an other issue which should be clarified: Obviously the covariant formalism without the modification leads to  CVE correctly \cite{glpww}, so that our modification can be seem to over count  the CVE. However,  this not the case. Within  the  modified QKE formalism of \cite{dk}  in 4D the first order solutions of chiral fields are different from the ordinary ones and they generate the CVE correctly as it was verified explicitly in \cite{dk}. When one integrates 4D CKE over the zeroth component of momentum to get the nonrelativistic CKT, the modification terms turn up to be essential in acquiring the CVE correctly. In fact,  as it will be explained at the end of  section  \ref{sec-cke}, the consistent 3D CKT which results from the  4D formalism would not generate the CVE correctly without the modification terms. }

We deal with relativistic plasma as a fluid whose  constituents are fermions obeying the Dirac equation. 
We will introduce the vector field $\eta_\mu (x)$ which is minimally coupled to Dirac fields as the electromagnetic vector potential $A_\mu(x).$ 
However,  it is not a $U(1)$ gauge field. Equations of motion of  the new vector  potential follow from an action which  describes fluid dynamics in terms of self interacting scalar field.  Although there are some crucial  differences, the action which we consider is mainly introduced  in \cite{bb}.   It is invariant under  a gauge transformation which is not the custom $U(1)$ symmetry.   
Dirac equation coupled to the new vector potential is also invariant under this gauge transformation when the Dirac field is transformed appropriately. 

To derive the equation of motion of the Wigner function one employs the equation satisfied by the Dirac field coupled to  vector potential and its gauge invariance  \cite{vge}. As far as the Dirac Lagrangian is considered, the form of the gauge transformations related to $\eta_\mu (x)$  and $A_\mu(x)$ are similar.  Hence, we  generalize  the procedure of \cite{vge} to derive the QKE satisfied the Wigner function  when both of the $\eta_\mu (x)$  and $A_\mu(x)$ gauge fields are present. Then, we  decompose the Wigner function in  the Clifford algebra basis and obtain  equations satisfied by the  fields which are the coefficients of Clifford algebra generators. These  equations depend explicitly on the field strengths. Some of these equations can be eliminated and the rest can be used to obtain kinetic  transport equations (KTE). In general, until acquiring KTE  electromagnetic field strength is defined in terms of vector potentials. Once the KTE are  established it is expressed by the electric and magnetic fields which satisfy the Maxwell equations. However, as it will be discussed,  for the field strength related to  fluid one may proceed in two different ways. The first option is to
require that the  field strength related to fluid is expressed in terms of vorticity and fluid velocity before deriving KTE. The second option is to  establish KTE first and then express the fluid related field strength in terms of vorticity and fluid velocity.  When the former method is adopted we find the KTE  proposed in \cite{dk,dk-m}.  The latter method which is similar to the electromagnetic case will be the subject of  this work. In this case the massless and  massive KTE   can be obtained by generalizing  the kinetic equations established, respectively, in \cite{hpy1,hsjlz} and \cite{hhy}.

We will acquire 3D kinetic equations by integrating the covariant equations over the zeroth component of four-momentum. For chiral fermions we will show that a novel 3D CKT is accomplished  in the presence of both external electromagnetic field and fluid vorticity which does not depend explicitly on the spatial coordinates.  Moreover, this theory possesses the  Coriolis force term and it is consistent with the chiral anomaly. It generates the chiral magnetic and vortical effects correctly.  When one deals with massive fermions kinetic equations of vector and axial-vector components of Wigner function depend on the  spin four-vector  $a^\mu$ \cite{hhy}.  We provide mass corrections to 3D chiral effects by  letting    $a^\mu$ be given with the free Dirac equation  for small mass values.

We start with presenting the  action which is considered to establish transport equations of a fermionic fluid interacting with electromagnetic fields. 
In section \ref{sec-fluid} we focus on the part of action  which we claim to govern  the dynamical evolution of  fermionic fluid.  
 In section \ref{sec-QKE} we present an outline of the derivation  of  QKE satisfied by the Wigner function of  Dirac fields coupled to two independent vector fields.  section \ref{sec-cke} is devoted to the study of the kinetic equations of chiral fermions in the external electromagnetic fields by taking into account noninertial properties of the fluid. The relativistic and the  3D  chiral transport equations are established. The massive fermions are studied in section \ref{sec-mass}, where the relativistic equations are integrated over the zeroth component of momentum  in the small mass limit by approximating the spin four-vector  adequately. Discussions of our results are presented in section \ref{sec-conc}.

\section{Action} 
\label{sec-Act}

To establish  quantum   kinetic equation  for   fermionic fluids in the presence of electromagnetic fields we propose  the action  
\begin{equation}
\label{Stot}
	S=	S_\ssD	+S_Q+S_{\ssE\ssM}+S_\zeta +S_\phi .
\end{equation}
The first term is the Dirac action,
\begin{equation}
	S_\ssD = \frac{1}{2}\int d^4x\,  \bar{\psi} \left(i\hbar \gamma^\mu \partial_\mu  -m\right)\psi .
\end{equation}
We consider two vector potentials coupled to Dirac fields. One of them is the $U(1)$ gauge field $A_\mu,$  
\begin{eqnarray}
	S_\ssQ & =& -Q\int d^4x \  \bar{\psi}  \gamma^\mu A_\mu  \psi ,
\end{eqnarray}
whose dynamical equations are generated by
\begin{eqnarray}
S_{\ssE\ssM}&=&-\frac{1}{4}\int d^4x \ F^{\mu \nu} F_{\mu\nu} . \label{SEM}
\end{eqnarray}
$Q$ is the electric charge and 
\begin{equation}
F_{\mu\nu}=\partial_{\mu}A_\nu -\partial_{\nu} A_\mu, \label{fmn}
\end{equation}
is the field strength of the  $U(1)$ gauge field which is
invariant under the gauge transformation
\begin{equation}
\label{Ag}
A_\mu (x) \rightarrow A_\mu (x) -\partial_\mu \Lambda (x).
\end{equation}
The other one is  the  real four-vector field  $\eta_ \alpha,$ whose coupling constant is $\zeta,$ 
\begin{eqnarray}
\label{szeta}
S_\zeta =  \zeta \int d^4x \bar{\psi}\gamma^\alpha\eta_\alpha \psi.
\end{eqnarray}
$\eta_\alpha$ is also  coupled   to the  complex scalar field $\phi,$  as follows,
\begin{equation}
\label{SPhi}
	S_\phi =\frac{1}{2} \int d^4x \left[\left(\partial^\alpha \phi -i\eta^\alpha\phi\right)^\star \left(\partial_\alpha \phi -i\eta_\alpha\phi\right)-V\left(\phi^\star \phi\right)\right].
\end{equation} 
In the next section we will clarify how the scalar field $\phi$ and the vector-field  $\eta_\alpha,$     represent the fluid.
We work in Minkowski spacetime with $g_{\mu\nu}=\diag (1,-1,-1,-1).$ 

\section{Fluid} 
\label{sec-fluid}

In this section we will describe  how the  variation of the action
\begin{equation}
\label{sF}
S_\ssF=S_\phi + S_\zeta  ,
\end{equation}
 with respect to  $\eta_\alpha$ and $ \phi $ fields generate effectively the Euler action of a fluid with Coriolis force, whose constituent particles  are  Dirac fermions. { The scalar field $\phi$ is the mean field which represents  fluid. The vector field $\eta_\alpha$ is an auxiliary field which will be fixed as in (\ref{main1}) below, hence it is considered only at the classical level.} Our formulation is mainly inspired from   \cite{bb}, where  a  covariant action was proposed to  describe  magnetohydrodynamics as a field theory. However, we are  interested in expressing the vector field $\eta_\alpha$ in terms of fluid variables like the enthalpy current. In contrary to \cite{bb}, in our formulation hydrodynamical  quantities are not considered as variables. They will be related to the independent field variables $\eta_\alpha,\phi$ by some other means as we will discuss. Therefore we do not need the constraints considered in  \cite{bb}.

Let us first write the complex scalar field in terms of two real fields: 
\begin{equation}
\label{phi}
\phi=\sigma e^{i\varphi}.
\end{equation}
Plugging this definition into  the action (\ref{SPhi}) yields 
\begin{equation}
\label{Sreal}
S_\phi =\frac{1}{2} \int d^4x \left[\partial^\alpha \sigma  \partial_\alpha \sigma+\sigma^2\left(\partial^\alpha \varphi -
\eta^\alpha\right) \left( \partial_\alpha \varphi -\eta_\alpha\right)-V(\sigma^2)\right].
\end{equation} 
Observe that it is invariant under the gauge transformation
\begin{equation}
\label{etag}
\eta_\alpha (x)\rightarrow \eta_\alpha (x) -\partial_\alpha \lambda (x),\ \ \ \varphi (x)\rightarrow \varphi (x)-\lambda (x).
\end{equation}

After  expressing $S_\phi $ as in  (\ref{Sreal}), variation of $S_\ssF$ with respect to  $\varphi$ generates the following equation of motion
\begin{equation}
\label{em2}
\partial_\alpha [\sigma^2\left(\partial^\alpha \varphi -\eta^\alpha\right)] =0.
\end{equation} 
On the other hand variation of $S_\ssF$ with respect to the $\eta_\alpha$ yields
\begin{equation}
\label{main0}
\sigma^2\left(\partial^\alpha \varphi -\eta^\alpha\right) -\zeta \bar{\psi}\gamma^\alpha \psi =0.
\end{equation} 

{ The mean value of particle number current density operator of the  Dirac particles can be  expressed   in the form
\begin{equation}
\label{pnc}
j^\alpha =\langle :\!\bar{\hat{\psi}} \gamma^\alpha \hat{\psi }\! :\rangle =\int d^4q\, q^\alpha f(x,q) .
\end{equation}
Here, $\hat{\psi},\ \bar{\hat{\psi}}$ are operators and  colons denote normal ordering.  The exact form of the distribution function 
$f(x.q)$ can be obtained from the Wigner function satisfying QKE  \cite{Groot}. 
  One can also describe this system as a fluid. For this purpose let us  introduce the fluid  four-velocity $u_\alpha\equiv dx_\alpha (\tau)/d\tau ,$ where $\tau$ is the proper time and $x_\alpha (\tau)$ is the  worldline of a fluid element, so that it satisfies
\begin{equation}
\label{u2}
u^\alpha u_\alpha=1.
\end{equation}
It can be used to  decompose the momentum four-vector as $q^\alpha=(u\cdot q)u^\alpha+ q_{\perp}^{\alpha},$ where 
$q_{\perp}\cdot u=0.$ Then, due to the momentum integral in (\ref{pnc}), one gets 
\begin{equation}
\label{nu1}
j^\alpha = nu^\alpha  ,
\end{equation}
where $n$ is  the particle number density. In principle due to the motion of  medium, particle number current density can have an anomalous parts. For example due to rotations there would be a term which depends linearly  on the vorticity of the medium.  
 However, we ignore the  anomalous contributions to current because we  consider only classical fields. 
In the  classical field (mean field) approximation the quantum field $\hat{\psi}$ can be considered as a collection of wave-packets constructed by the solutions of  Dirac equation.  Then, when we deal with fluids composed of spin-1/2 particles,   a fluid element which contains a large  number of particles but compact enough such that they behave  homogeneously, can be taken to coincide with 
one of the  wave-packets. 
Therefore,  in the  mean field approximation we can write 
\begin{equation}
\label{nu}
\bar{\psi}\gamma^\alpha \psi \equiv nu^\alpha  .
\end{equation}}
Now  (\ref{main0})  can equivalently be written as
\begin{equation}
\label{main}
nu^\alpha =\zeta^{-1} \sigma^2\left(\partial^\alpha \varphi -\eta^\alpha\right) ,
\end{equation} 
and (\ref{em2})  states that  the particle number current density is conserved:
\begin{equation}
\label{em21}
\partial_\alpha \left(nu^\alpha\right) =0.
\end{equation} 
In fact, in  relativistic fluid dynamics particle number current density without dissipation is given  with  (\ref{nu1})  \cite{LanLif}.

Equipped with these relations we may now discuss why we consider the action (\ref{Stot}). The kinetic theory of a neutral plasma can be described in terms of the
relativistic scalar electrodynamics with two
 scalar fields $\phi_1$ and $\phi_2$ which are macroscopic wave functions representing the negative and positive charge carriers \cite{fa-ni,ffnr}:
\begin{equation}
S_{np}=\int d^4x \left[|\partial_\alpha \phi_1 +iQA_\alpha\phi_1|^2 +|\partial_\alpha \phi_2 -iQA_\alpha\phi_2|^2-V\left(\phi_1, \phi_2\right)\right]+S_{\ssE\ssM}.
\end{equation}
If one would like to represent neutral plasma in terms of one scalar field $\phi$ instead of $\phi_1$ and $\phi_2,$ it cannot minimally couple to $A_\mu.$ Then, it should  interact with electromagnetic fields in a complicated way \cite{bb}. In fact (\ref{main0}) represents this interaction because the variation of the action (\ref{Stot}) with respect to $A_\mu$  shows that (\ref{pnc}) is the  current which  describes how the electromagnetic fields will evolve in plasma:
\begin{equation}
\label{r21}
\partial^\mu F_{\mu\nu}= Q\bar{\psi}\gamma_\nu \psi .
\end{equation}
By inspecting (\ref{main}) we see that interaction of the fluid with electromagnetic fields is exposed by setting 
\begin{equation}
\label{r22}
 \eta^\alpha= \partial^\alpha \varphi -\frac{\zeta}{Q\sigma^2}\partial_\mu F^{\mu\alpha}.
\end{equation}
 In fact by inserting (\ref{r21}) and (\ref{r22}) into the action (\ref{sF}) we get
\begin{equation}
\label{sFr}
S_\ssF= \int d^4x \Big\{\frac{1}{2} \left[\partial^\alpha \sigma  \partial_\alpha \sigma-V(\sigma^2) \right]
 +\frac{\zeta}{Q}(\partial_\mu F^{\mu\alpha})\partial_{\alpha }\varphi-\frac{\zeta^2}{2Q^2\sigma^2} (\partial_\mu F^{\mu\alpha})(\partial^\nu F_{\nu\alpha})
\Big\}.
\end{equation}
This shows that scalar field components  couple to the derivatives of the electromagnetic fields which would be calculated from the Maxwell equations (\ref{r21}) whose charge and current  distributions are due to the charged fermions. Hence, the interactions between the scalar field and electromagnetic fields are sensitive how the electric and magnetic fields change spatially and temporally.

Variation of $S_\ssF$ with respect to $\sigma$ leads to
\begin{equation}
\label{em3}
\partial^\alpha\partial_\alpha\sigma+\sigma\left(\partial^\alpha \varphi -
\eta^\alpha\right) \left( \partial_\alpha \varphi -\eta_\alpha\right) -\sigma V^\prime (\sigma^2)=0,
\end{equation}
where $V^\prime$ 
is the derivative of $V$ with respect to its argument. As in \cite{bb} we assume that amplitude of $\phi$  varies slowly so that the first term is neglected compared to the second one. Thus we get 
\begin{equation}
\label{eqm2}
\left(\partial^\alpha \varphi -
\eta^\alpha\right) \left( \partial_\alpha \varphi -\eta_\alpha\right) =\nu^2,
\end{equation}
where we defined
$$
\nu^2 \equiv V^\prime (\sigma^2).
$$
Using  (\ref{main}) in (\ref{eqm2}) yields 
\begin{equation}
n^2\zeta^2=\sigma^4\nu^2,
\end{equation}
so that one can express $n$ as a function of the real scalar field $\sigma$ as 
\begin{equation}
\label{ssoz}
n=\frac{\sigma^2 \nu}{\zeta}.
\end{equation}
Plugging (\ref{ssoz}) back into (\ref{main}) leads to
\begin{equation}
\label{main1}
\partial^\alpha \varphi -\eta^\alpha = \nu u^\alpha .
\end{equation} 
By taking the derivative of (\ref{eqm2}) and employing (\ref{main1}), we attain
\begin{equation}
\label{2der}
 \nu u^\alpha  (\partial_\beta \partial_\alpha \varphi -\partial_\beta \eta_\alpha )=\nu \partial_\beta \nu.
\end{equation} 
Let us write it as
\begin{equation}
\label{22der}
\nu u^\alpha  (\partial_\alpha\partial_\beta \varphi -\partial_\alpha\eta_\beta -W_{\alpha \beta} - w_{\beta \alpha} )=\nu \partial_\beta \nu,
\end{equation} 
where we  introduced
\begin{equation}
w_{\beta \alpha}=\partial_\beta\eta_\alpha -\partial_\alpha \eta_\beta,
\label{wmn}
\end{equation}
and
$$
W_{\alpha \beta}= \partial_{\alpha} \partial_{\beta } \varphi - \partial_{\beta} \partial_{\alpha } \varphi .
$$
Obviously $W_{\alpha \beta}$ vanishes for ordinary functions but in 
it can also be different from zero \cite{bb}:  $\varphi$  is the phase of  scalar field $\phi,$  (\ref{phi}),  so that it is defined up to some functions $\theta (x)=(\theta_1 (x),\theta_2 (x), \cdots)$ satisfying
$$
e^{\theta_k (x)}=1.
$$
Hence, along some $\theta_k (x)=0,$ curves, the mixed partial derivatives of $\varphi$ can fail to be continuous.  We can ignore the singular curves and set $W_{\alpha \beta}=0$ without loss of generality. Nevertheless, we  can also consider  singular cases where the following condition is satisfied,
\begin{equation}
\label{W0}
u^\alpha W_{\alpha \beta}=0.
\end{equation}
Now by taking the derivative of (\ref{main1}),
$$
\partial_\alpha\partial_\beta \varphi -\partial_\alpha\eta_\beta=u_\beta \partial_\alpha \nu +\nu \partial_\alpha u_\beta ,
$$
and using it in (\ref{22der}) we obtain
\begin{equation}
\label{23der}
\nu u^\alpha  (u_\beta \partial_\alpha \nu+\nu \partial_\alpha u_\beta  - w_{\beta \alpha} )=\nu \partial_\beta \nu .
\end{equation} 
As far as $\nu \neq 0,$ it yields
\begin{equation}
\label{24der}
 \nu u^\alpha  \partial_\alpha u_\beta = \partial_\beta \nu -u^\alpha u_\beta \partial_\alpha \nu+u^\alpha  w_{\beta \alpha} .
\end{equation} 
Acceleration which is defined as the derivative of the velocity $u_\alpha$ with respect to the proper time $\tau ,$ can be calculated by making use of (\ref{24der}):
\begin{equation}
\label{25der}
\frac{d u_\beta}{d \tau}= u^\alpha  \partial_\alpha u_\beta = \frac{\partial_\beta \nu}{\nu } -u^\alpha u_\beta \frac{\partial_\alpha \nu}{\nu}+
\frac{u^\alpha   w_{\beta \alpha} }{\nu}.
\end{equation} 

{   Let us compare (\ref{25der}) with the (hydrodynamic) Euler equations 
\begin{equation}
\label{euler}
\frac{d u_\beta}{d \tau}= \frac{\partial_\beta P}{\rho +P } -u^\alpha u_\beta \frac{\partial_\alpha P}{\rho +P}-\frac{F_{\beta }}{\rho +P},
\end{equation} 
where $\rho$ is  the energy density, $P$ is the pressure. They are related to the  specific enthalpy $h$ as $nh=\rho +P.$ Here $F_\alpha$ is an external force
 which can be the gravitational force, electromagnetic force and the ``fictitious'' force such as the centrifugal or Coriolis force \cite{ReZo}.  

Except their last terms, (\ref{25der}) and  (\ref{euler})  are identical if 
\begin{equation}
\label{equ}
\frac{dP}{\rho +P}=\frac{d\nu}{\nu}.
\end{equation}
{From the first law of thermodynamics one knows that
\begin{equation}
\label{the}
\frac{d\rho}{dn}=\frac{\rho +P}{n},
\end{equation}
for the  fluids of only one kind of particle whose total  number is conserved and without heat exchange.  }
Suppose that (\ref{equ}) is satisfied, then by making use of (\ref{the}) we have
\begin{equation}
\label{nls}
\frac{d (\rho +P)}{dn}=\frac{\rho +P}{n}+\frac{dP}{d\nu}\frac{d\nu}{dn}=\frac{\rho +P}{n}+\frac{\rho +P}{\nu} \frac{d\nu}{dn}.
\end{equation}
It can be expressed as 
\begin{equation}
\label{nls1}
\frac{d (\rho +P)}{\rho +P}=\frac{dn}{n}+ \frac{d\nu}{\nu}.
\end{equation}
By integrating it 
\begin{equation}
\label{nuent}
\nu=\xi\frac{\rho +P}{n}=\xi h
\end{equation}
follows, where $\xi$ is a positive constant of integration.

We need to express $\nu$ analytically in terms of fluid's parameters by  respecting the equality (\ref{nuent}).  Let us consider the equation of state $P=(\gamma -1)\rho ,$ where $\gamma>1$ is the adiabatic index. This relation can be taken as the definition of
an ideal fluid  \cite{ReZo} and it is consistent with the equation of state resulting from the field equations by choosing $V(\sigma^2)$ adequately \cite{bb}. Then, we can write
\begin{equation}
\label{nhr}
h= \gamma \frac{\rho }{n}.
\end{equation}
By plugging it into  (\ref{nuent}) we get
\begin{equation}
\nu =\xi^\prime \frac{\rho }{n},
\end{equation}
with  $\xi^\prime =\xi \gamma.$ In the mean field approximation we identified  a fluid element with a wave packet.  Hence the proper energy density can be parametrized in terms of the momentum $p^\mu$ of the wave packet center as
\begin{equation}
\frac{\rho }{n}=  u\cdot p.
\end{equation}
Therefore,  we write 
\begin{equation}
\label{nup}
\nu =\xi^\prime u\cdot p.
\end{equation}

From  (\ref{main1}) by setting $W_{\alpha \beta}=0$ or employing the condition  (\ref{W0}), we get
\begin{equation}
\label{aaa}
u^\alpha   w_{ \alpha \beta} =u^\alpha \left[ \partial_\beta(\nu u_\alpha)- \partial_\alpha(\nu u_\beta)\right] \equiv \xi^\prime  {\cal K}_\beta.
\end{equation}
Let us demonstrate that  ${\cal K}_\beta \equiv u^\alpha\left[ \partial_\beta(u\cdot p\,  u_\alpha)- \partial_\alpha(u\cdot p\,  u_\beta)\right] , $ is the relativistic generalization of Coriolis force per particle.
We consider vanishing  linear acceleration 
\begin{equation}
\label{vanac}
u^\alpha \partial_\alpha u_\beta=0,
\end{equation}
so that the fluid velocity can be taken to satisfy
\begin{equation}
\partial_\alpha u_\beta =-\partial_\beta u_\alpha .
\end{equation}
Thus we write
$$
{\cal K}_\beta = p^\alpha\partial_\beta u_\alpha .
$$
Fluid vorticity four-vector   is defined as 
\begin{equation}
\omega^\mu =\frac{1}{2} \epsilon^{ \mu \nu \alpha \beta}\Omega_{\alpha \beta}u_\nu,
\end{equation}
where 
\begin{equation}
\Omega_{\alpha \beta}= \frac{1}{2}(\partial_\alpha u_\beta -\partial_\beta u_\alpha ) ,
\end{equation}
is the kinematic vorticity tensor.
Therefore we get
$$
{\cal K}_\beta =\epsilon_{ \beta \alpha \mu \nu}u^\mu \omega^\nu p^\alpha.
$$
In the frame  $u^\alpha=(1,\bm 0),\ \omega^\alpha =(0, \bm \omega),$ it becomes ${\cal K}_\beta =(0, \bm{{\cal K}})$ with 
$$
\bm{{\cal K}}=\bm p \times \bm \omega.
$$
Hence, we conclude that the last terms of  (\ref{25der}) and (\ref{euler}) are identical where 
$$F_\alpha = n\gamma\, {\cal K}_\alpha ,$$ 
is a relativistic extension of the Coriolis force  in a rotating coordinate frame.

From (\ref{aaa})  and (\ref{nup}) we have
\begin{equation}
\label{wab}
w_{ \alpha \beta} =\xi^\prime \left[ (\partial_\beta u\cdot p) u_\alpha-(\partial_\alpha u\cdot p) u_\beta \right] +\kappa\, u\cdot p \left(\partial_\beta u_\alpha- \partial_\alpha u_\beta \right).
\end{equation}
On the other hand, due to the vanishing of linear acceleration, (\ref{vanac}),  we observe that
$$
u^\alpha \left(\partial_\beta u_\alpha- \partial_\alpha u_\beta \right)=0.
$$
Hence, (\ref{aaa}) is satisfied for  an arbitrary constant $\kappa,$  which can be even zero.
By introducing 
\begin{eqnarray}
\label{wmuC}
w^{\mu\nu}_\ssC&= &(\partial^\mu u^\alpha )  p_\alpha  u^\nu - (\partial^\nu u^\alpha )  p_\alpha  u^\mu, 
\end{eqnarray}
 $w_{\mu\nu}$ can be expressed as
\begin{equation}
\label{wab1}
w^{\mu\nu}=\xi^\prime w^{\mu\nu}_\ssC +\kappa (u\cdot p) \Omega^{\mu\nu}.
\end{equation}
It is worth noting that  $\xi,\ \kappa$  are arbitrary constants and   $w^{\mu\nu}$ is the circulation (vorticity) tensor  for $  \kappa=2\gamma ,$
and $ \xi=1,$ i.e. $\xi^\prime =\gamma ,$ \cite{ReZo}.
Therefore, we come to the conclusion that the scalar field $\phi$ and the vector field $\eta_\alpha$ represent the fluid composed of the Dirac particles. Moreover,  the field strength of $\eta_\alpha$ is given by (\ref{wab1}) when  the equations of motion resulting from the variation of $S_\ssF$  with respect to $\phi$ and $\eta_\alpha$ are satisfied.}

\section{Quantum Kinetic Equation}
\label{sec-QKE}

Let us return to the initial action (\ref{Stot})  without imposing the equations of motion derived from  (\ref{SEM}) and  (\ref{sF}). Now, we would like to examine  the action of  Dirac field  coupled to the vector fields,
\begin{equation}
\label{DAE}
S_\psi\equiv 	S_\ssD	+S_Q+S_\zeta = \frac{1}{2}\int d^4x\,  \bar{\psi} [ \gamma^\mu (i\hbar \partial_\mu - \zeta \eta_\mu -QA_\mu ) -m] \psi .
\end{equation}
It generates the Dirac  equation 
\begin{equation}
\label{deq}
[ \gamma^\mu (i\hbar \partial_\mu - \zeta \eta_\mu -QA_\mu ) -m] \psi =0,
\end{equation}
which is invariant under the gauge transformations   (\ref{Ag}) and  (\ref{etag}), when the spinor field transforms  as
\begin{eqnarray}
\psi (x)\rightarrow e^{i (\zeta\lambda (x) +Q\Lambda (x) )/ \hbar}\psi (x).  \label{G1}
\end{eqnarray}

The Wigner operator is defined by
\begin{equation}
\label{wo1}
\hat{W}(x,p) = \int d^4y\,  e^{-ip\cdot y/\hbar}  \bar{\psi}(x) e^{\frac{1}{2}y\cdot\partial^\dagger}\otimes e^{-\frac{1}{2}y\cdot\partial} \psi(x).
\end{equation}
Here, $\psi (x)$ and $\bar{\psi} (x)$ are operators and $\otimes$ represents tensor product. 
The Wigner function is defined as the ensemble average of the normal ordered Wigner operator:
\begin{equation}
\label{wfun}
W(x,p)=\langle :\!\hat{W}(x,p)\!: \rangle .
\end{equation}

To derive the kinetic equation satisfied by the Wigner function we  proceed as it was done in \cite{vge}:
The  Wigner operator defined in  (\ref{wo1}) fails to be  invariant under the gauge transformations   (\ref{Ag}), (\ref{etag}) and (\ref{G1}). To define the gauge invariant Wigner operator  one introduces the gauge link
\begin{equation}
\label{link}
U(A,\eta; x_1,x_2)\equiv \exp \left[-iQ\gamma^\mu \int_0^1 ds A_\mu (x_2+sy)\right] \exp \left[-i \zeta \gamma^\mu \int_0^1 ds  \eta_\mu (x_2+sy)\right],
\end{equation}
where $x_1^\mu\equiv x^\mu+y^\mu /2,$ $x_2^\mu\equiv x^\mu-y^\mu /2,$ and  insert it into (\ref{wo1}):
\begin{equation}
\label{wo2}
\hat{W}(x,p)=\int d^4y\, e^{-ip\cdot y/\hbar}   \bar{\psi}(x_1)  U(A,\eta; x_1,x_2) \otimes \psi (x_2).
\end{equation}
By making use of the Dirac equation (\ref{deq}), one can show that the Wigner operator, (\ref{wo2}), satisfies 
\begin{eqnarray}
[  \gamma \cdot (p-\frac{1}{2}i\hbar \partial ) -m ] \hat{W}(x,p)=-i\hbar\partial_p^\mu 
\int d^4y\,  e^{-ip\cdot y/\hbar}   \bar{\psi}(x_1) \otimes
 \nonumber\\
\Big[ \int_{0}^{1} ds (1-s)
\Big(QF_{\mu\nu}(x+sy-y/2)+ \zeta w_{\mu\nu} (x+sy-y/2) \Big) \label{qeq}\\
 U(A,\eta; x_1,x_2)\Big]\gamma_\nu\psi (x_2), \nonumber
\end{eqnarray}
where $\partial_p^\mu \equiv \partial / \partial p_\mu .$ $F_{\mu\nu}$ and  $w_{\mu\nu}$ are  defined by (\ref{fmn}) and (\ref{wmn}). We consider the mean field approximation, so that the field strengths $F_{\mu\nu},\ w_{\mu\nu} $ are c-valued fields. Following \cite{vge} we write
$$
f(x+sy-y/2) ]=e^{(s-1/2)y\cdot\partial }f(x)
$$
and employ the relation
$$
\int d^4y\, e^{-ip\cdot y/\hbar}f(y)g(y)=f(i\hbar \partial_p) \int d^4y\,  e^{-ip\cdot y/\hbar}g(y),
$$ 
to express the right-hand-side of (\ref{qeq}) as
\begin{eqnarray}
&i\hbar \int_{0}^{1} ds (s-1)e^{(s-1/2)i\hbar \partial_p\cdot \partial}
\Big(QF_{\mu\nu}(x)+ \zeta w_{\mu\nu} (x) \Big)\partial_p^\mu 
\int d^4y\,   e^{-ip\cdot y/\hbar}   \bar{\psi}(x_1) \otimes U(A,\eta; x_1,x_2)\gamma_\nu\psi (x_2) .\nonumber
\end{eqnarray}
One  expands the exponential as power series and perform the $s$ integration. Then
by taking the ensemble average of (\ref{qeq}), 
the quantum kinetic equation satisfied by the Wigner function is established as
\begin{equation}
\left[\gamma \cdot \left(\pi+ \frac{i\hbar}{2} D \right)-m \right] W(x,p) = 0,
\label{qke}
\end{equation}
 with
\begin{eqnarray}
D^{\mu} &\equiv & \partial^{\mu}-j_{0}(\Delta)\left[  QF^{\mu\nu}  + \zeta w^{\mu\nu} \right]  \partial_{p \nu} , \label{Df}\nonumber\\
\pi^{\mu} &\equiv &p^{\mu}-\frac{\hbar}{2} j_{1}(\Delta) \left[ Q F^{\mu\nu}  +\zeta w^{\mu\nu} \right]  \partial_{p \nu} .\label{Pf} \nonumber
\end{eqnarray}
\(j_{0}(x)\) and \(j_{1}(x)\)
are spherical Bessel functions in  \(\Delta \equiv \frac{\hbar}{2} \partial_{p} \cdot \partial_{x}\).  The space-time derivative \(\partial_{\mu}\) contained in \(\Delta\) acts on \(\left[ Q F^{\mu\nu}  + \zeta w^{\mu\nu} \right] ,\) but not on the Wigner function. In contrary  $ \partial_{p}^{\nu}$ acts on the Wigner function, but not on  \(\left[ Q F^{\mu\nu}  +\zeta w^{\mu\nu} \right] .\)  

The  Wigner function can be decomposed by means of the 16 generators of the Clifford algebra 
\begin{equation}
W=\frac{1}{4}\left(\mathcal{F}+i \gamma^{5} \mathcal{P}+\gamma^{\mu} \mathcal{V}_{\mu}+\gamma^{5} \gamma^{\mu} \mathcal{A}_{\mu}+\frac{1}{2} \sigma^{\mu \nu} \mathcal{S}_{\mu \nu}\right),
\label{wigner}
\end{equation}
where the coefficients $\mathcal{C}\equiv \{\mathcal{F},\mathcal{P},\mathcal{V}_{\mu},\mathcal{A}_{\mu},\mathcal{S}_{\mu \nu}\},$  respectively, are the scalar, pseudoscalar, vector, axial-vector, and antisymmetric tensor fields. 
We expand them in powers of Planck constant and keep the leading and next to the leading order terms in $\hbar.$ Hence in (\ref{qke}) 
one sets  $\pi_\mu=p_\mu$   and substitutes  $D_\mu$ with 
\begin{eqnarray}
\tilde{\nabla}^\nu &\equiv  & \partial^{\nu}-\left[  QF^{\nu\beta} +\zeta  w^{\nu\beta} \right]  \partial_{p \beta} .\end{eqnarray}
Plugging   (\ref{wigner}) into  (\ref{qke}),  yields complex valued equations 
whose  real parts are
\begin{eqnarray}
p\cdot \mathcal{V}-m \mathcal{F} =0,  \label{real1} \\
{p_{\mu} \mathcal{F}-\frac{\hbar}{2} \tilde{\nabla}^{\nu} \mathcal{S}_{\nu \mu}-m \mathcal{V}_{\mu}=0},  \label{real2} \\
{-\frac{\hbar}{2} \tilde{\nabla}_{\mu} \mathcal{P}+\frac{1}{2} \epsilon_{\mu \nu \alpha \beta} p^{\nu} S^{\alpha \beta}+m \mathcal{A}_{\mu}=0}, \label{real3} \\
{\frac{\hbar}{2} \tilde{\nabla}_{[\mu} \mathcal{V}_{\nu]}-\epsilon_{\mu \nu \alpha \beta} p^{\alpha} \mathcal{A}^{\beta}-m \mathcal{S}_{\mu \nu}=0}, \label{real4} \\
\frac{\hbar}{2} \tilde{\nabla} \cdot \mathcal{A}+m \mathcal{P} =0, 
\label{real5}
\end{eqnarray}
and the imaginary parts are
\begin{eqnarray}
{\hbar \tilde{\nabla} \cdot \mathcal{V}=0}, \label{imag1} \\ 
{p\cdot \mathcal{A}=0}, \label{imag2} \\ 
{\frac{\hbar}{2} \tilde{\nabla}_{\mu} \mathcal{F}+p^{\nu} \mathcal{S}_{\nu \mu}=0},  \label{imag3} \\ 
{p_{\mu} \mathcal{P}+\frac{\hbar}{4} \epsilon_{\mu \nu \alpha \beta} \tilde{\nabla}^{\nu} \mathcal{S}^{\alpha \beta}=0},  \label{imag4} \\ 
{p_{[\mu} \mathcal{V}_{\nu]}+\frac{\hbar}{2} \epsilon_{\mu \nu \alpha \beta} \tilde{\nabla}^{\alpha} \mathcal{A}^{\beta}=0}.
\label{imag5}
\end{eqnarray}

To derive QKE (\ref{qke}), we started from the Dirac equation (\ref{DAE}). Then in  getting (\ref{qeq})  we had to  introduce $F_{\mu\nu}$ and $w_{\mu\nu}$ 
which are defined in terms of the gauge fields as in (\ref{fmn}) and (\ref{wmn}).  
The equations of motions  following from the action (\ref{SEM}) give  the  Maxwell equations when 
one can expresses the field strength in terms of   the electromagnetic fields $E_\mu, B_\mu$ by
\begin{equation}
\label{fEM}
F^{\mu\nu}=E^\mu u^\nu -E^\nu u^\mu +\epsilon^\mnar u_\alpha B_\rho,
\end{equation}
where $u_\mu$ is the fluid 4-velocity. Similarly, when the equations of motion discussed in section \ref{sec-fluid} are imposed,  one deals with the Euler equations (\ref{euler}), where $w_{\mu\nu}$  is expressed in terms of vorticity and {energy per particle}  as in (\ref{wab1}). 
However, when the equations of motion following from (\ref{SEM}) and (\ref{sF}) should be imposed? We have two options:  $i)$ Obtain the kinetic equations which the field components of Wigner function, $\mathcal{C},$ satisfy and then impose the equations of motion.
$ii)$	Impose the equations of motion  from the beginning and then  derive the kinetic equations satisfied by  the fields $\mathcal{C}.$ Our previous works \cite{dk,dk-m}  are consistent with the latter option for $\eta_\mu.$  Although we kept $F_{\mu\nu}$ as in (\ref{fmn}),  $w_{\mu\nu}$  was expressed in terms of enthalpy current,  (\ref{wab}), and then derived the kinetic equations satisfied by the fields $\mathcal{C}.$ Here, we   adopt the former option for both of them: We use the definition of $w_{\mu\nu}$ in terms of the $\eta_\mu$ fields, (\ref{wmn}), to establish  the kinetic equations and then apply the equations of motion yielding (\ref{wab}). Similarly, $F_{\mu\nu}$ is given by (\ref{fmn}) until the KTE are derived. Afterwards, we express it in terms of the electromagnetic fields, (\ref{fEM}). Once we choose  this method the semiclassical kinetic equations can directly be read from the known ones \cite{hpy1,hpy2,hsjlz,hhy}, by substituting $QF_{\mu\nu}$ with $QF_{\mu\nu}+\zeta w_{\mu\nu},$ as it will discussed in the subsequent sections.

\section{Chiral Kinetic Equations}
\label{sec-cke}
For vanishing mass the equations of $\mathcal{A}_\mu$ and  $\mathcal{V}_\mu$ decouple from the rest. They are given by  (\ref{real1}),(\ref{real4}),(\ref{real5}) with $m=0$ and (\ref{imag1}),(\ref{imag2}),(\ref{imag5}).
Let us  introduce  the chiral vector fields 
\begin{equation}
{\cal J}^\mu_\sschi = \frac{1}{2} ({\cal V}^\mu + \chi {\cal A}^\mu), \nonumber
\end{equation} 
where $\chi =1,$  and $\chi =-1,$ correspond to the right-handed and the left-handed  fermions.
They need to satisfy 
\begin{eqnarray}
p_\mu {\cal J}_\sschi^\mu & = & 0,
\label{1st,0} \\
\tilde{\nabla}^\mu {\cal J}_{\sschi \mu }& = & 0,
\label{2nd,0}\\
\hbar \epsilon_{\mu \nu \alpha \rho} \tilde{\nabla}^\alpha {\cal J}^\rho_\sschi&=& - 2 \chi (p_\mu {\cal J}_{\sschi \nu} -  p_\nu {\cal J}_{\sschi\mu}) .
\label{third,0}
\end{eqnarray} 
 The chiral kinetic equation which results from (\ref{1st,0})-(\ref{third,0}) can be acquired by generalizing the formalism given in  \cite{hpy1,hpy2,hsjlz}.
First, one can verify that the solution of  (\ref{1st,0}) and (\ref{third,0}) is 
\begin{eqnarray}
{\cal J}^{\mu}_{\sschi} &=& p^\mu f_\sschi \delta(p^2) + \frac{\hbar}{2}\chi  \epsilon^{\mu \nu \alpha \beta}\left(QF_{\alpha \beta} +\zeta w_{\alpha \beta}\right)  p_\nu f^{0}_{\sschi} \delta^\prime (p^2)  \nonumber\\
&+& \hbar \chi    S^{\mu \nu}_\ssbfn (\tilde{\nabla}_{ \nu} f^{0}_{\sschi}) \delta(p^2),   
\label{generalform}
\end{eqnarray}
where $ \delta^\prime (p^2) = - \delta(p^2)/p^2$ and  $f_\chi (x,p)\equiv  f^{0}_\sschi (x,p)+\hbar  f^{1}_\sschi (x,p)$ is  the   distribution function.
$n^\mu$ is an arbitrary vector satisfying $n^2=1$ and 
$$
S^{\mu \nu}_\ssbfn=
\frac{1}{ 2 n \cdot p}  \epsilon^{\mu \nu \rho \sigma} p_\rho  n_\sigma  .
$$ 
Then, by inserting (\ref{generalform}) into (\ref{2nd,0})  one acquires the chiral kinetic equation:
\begin{eqnarray}
&& \delta\left( p^2 + \hbar \chi Q \frac{n_\mu \tilde{F}^{\ssmn}   p_\nu }{n \cdot p}+ \hbar \chi \zeta \frac{n_\mu \tilde{w}^{\ssmn}   p_\nu}{n \cdot p} \right)   
\{ p \cdot \tilde{\nabla}  \nonumber\\
&& + \frac{\hbar \chi Q }{n \cdot p} S^{\mu \nu}_\ssbfn \ F_{\mu\rho}n^\rho  \tilde{\nabla}_\nu  
- \frac{\hbar \chi}{2n \cdot p}  \epsilon^{\mu \nu \lambda \rho } (\partial_\mu n_\nu )p_\lambda  \tilde{\nabla}_\rho
\label{CKTn}\\
&&
+\frac{\hbar \chi \zeta }{n \cdot p} S^{\mu \nu}_\ssbfn  w_{ \mu\rho} n^\rho\tilde{\nabla}_\nu 
-\frac{\hbar \chi  }{n \cdot p} S^{\mu \nu}_\ssbfn (\partial_\mu n_\alpha)p^\alpha\tilde{\nabla}_\nu  
\} f_\chi =0,\nonumber
\end{eqnarray}
where  $\tilde{F}_{\ssmn}=\frac{1}{2}\epsilon_{\mu \nu \alpha \rho}F^{\alpha \rho}$ and
$\tilde{w}_{\ssmn}=\frac{1}{2}\epsilon_{\mu \nu \alpha \rho}w^{\alpha \rho},$ are the dual field strengths.

Until now we have  $w_{\mu\nu}=\partial_\mu\eta_\nu - \partial_\nu\eta_\mu, $ because  $\eta_\mu$ was off-shell.
Now, we impose the equations of motion (\ref{main})-(\ref{em3}), thus $w_{\mu\nu}$ is given by (\ref{wab1}). In the rest frame of massive fermions { energy per particle is $m.$} Therefore, in the massless limit  we set $\kappa=0,$ and  write 
\begin{equation}
\zeta w^{\mu\nu}|_{m=o}\equiv k w^{\mu\nu}_\ssC =k\left[(\partial^\mu u^\alpha )  p_\alpha  u^\nu - (\partial^\nu u^\alpha )  p_\alpha  u^\mu\right],
\end{equation}
where  $k=-\zeta\kappa $ is an arbitrary constant which will be fixed.  Now, (\ref{CKTn}) is the chiral kinetic equation where the vorticity and electromagnetic tensors are treated on the same footing.

\subsection{3D Chiral kinetic theory}
To establish  the 3D chiral kinetic theory we write  the covariant transport equation  (\ref{CKTn}) in the comoving frame by setting $n^\mu=u^\mu,$   
\begin{eqnarray}
&& \delta\left( p^2 + \hbar \chi Q \frac{u_\mu \tilde{F}^{\ssmn}   p_\nu }{u \cdot p}+k \hbar \chi u_\mu \tilde{w}^{\ssmn}_\ssC   p_\nu \right)   
\{ p \cdot \tilde{\nabla}  \nonumber\\
&& + \frac{\hbar \chi Q }{u \cdot p} S^{\mu \nu}F_{\mu \rho}u^\rho \tilde{\nabla}_\nu  
 - \frac{\hbar \chi}{u \cdot p}  p_\mu \tilde{\Omega}^{\ssmn} \tilde{\nabla}_\nu 
\label{cke1}\\
&&
+k \frac{\hbar \chi }{u \cdot p} S^{\mu \nu} w_{\ssC  \mu\beta} u^\beta\tilde{\nabla}_\nu 
-\frac{\hbar \chi  }{u \cdot p} S^{\mu \nu}(\partial_\mu u_\alpha)p^\alpha\tilde{\nabla}_\nu  
  \} f_\chi =0. \nonumber
\end{eqnarray}

We defined  $\tilde{w}_{\ssC \mu \nu }=\frac{1}{2}\epsilon_{\mu \nu \alpha \rho} w^{\alpha \rho}_\ssC$ and 
\begin{equation}
\label{smunu}
S^{\mu \nu}=
\frac{1}{ 2 u \cdot p}  \epsilon^{\mu \nu \rho \sigma} p_\rho  u_\sigma  .
\end{equation}
Obviously,
$u_\mu \tilde{w}_\ssC^{\ssmn} =0,$ and due to the fact that linear acceleration vanishes (\ref{vanac}),  we have
$$
w_{\ssC \mu \beta }u^\beta =(\partial_\mu u_\alpha)p^\alpha .
$$
Thus, (\ref{cke1}) can be written as
\begin{equation}
\delta\left( p^2 - \hbar \chi Q\frac{ p \cdot B}{u\cdot p}\right)   
\{ p \cdot \tilde{\nabla}   + \frac{\hbar \chi Q }{u \cdot p} S^{\mu \nu}E_{\mu}  \tilde{\nabla}_\nu  
- \frac{\hbar \chi}{u \cdot p}  p_\mu \tilde{\Omega}^{\ssmn} \tilde{\nabla}_\nu
+(k-1)\frac{\hbar \chi  }{u \cdot p} S^{\mu \nu} \Omega_{\mu \alpha}p^\alpha\tilde{\nabla}_\nu  
\} f_\chi =0,  \label{cke12}
\end{equation}
where  $E_\mu$ and $B_\mu$  are  the external electromagnetic fields. { Kinetic vorticity tensor $\Omega_{\mu \nu}$ can be expressed as  
\begin{equation}
\label{Ommunu}
\Omega_{\mu \nu}=\epsilon_{\mu \nu\alpha \beta }u^\alpha \omega^\beta
\end{equation}}
and its dual is ${\tilde{\Omega}}^{\mu\nu}=\omega_\mu u_\nu -u_\mu \omega_\nu . $


To establish a 3D CKT  we  would like to integrate (\ref{cke12}) over $p_0.$ To perform this integration we decompose 
 the distribution function into particle $s=1$ and antiparticle $s=-1$ parts,
\begin{equation}
f_{\sschi} (x,p)= \sum_{s=\pm 1} \theta(s  u\cdot p) f^{s}_{\sschi} (x,p).
\label{f_fd}
\end{equation}
Moreover, we choose   the frame  
\begin{equation}
\label{cuo}
u^\alpha=(1,\bm 0)\ \ \mathrm{and}\ \   \omega^\alpha =(0, \bm \omega),
\end{equation} 
where  the
 delta function yields the dispersion relations
\begin{equation}
{\cal E}^\sschi_s=s|\bm p|\left( 1 -\hbar \chi Q\frac{\bm B \cdot \bm p}{2|\bm p|^3}\right).
\end{equation}
Let us also introduce the canonical velocity
\begin{equation}
\label{canvel}
\bm v^\sschi_s\equiv \frac{\partial {\cal E}^\sschi_s}{\partial \bm p}=s \hat{\mathbf p}(1+s\hbar    \chi Q \frac{\bm B \cdot \bm p}{|\bm p|^3})- \hbar    \chi Q \frac{\bm B}{2|\bm p|^2}.
\end{equation}

As we show in appendix  \ref{appe}, integrating (\ref{cke12}) over $p_0$ leads to the transport equation
\begin{equation}
\Big( \sqrt{\eta}^{\, \sschi }_s \frac{\partial }{\partial t} + (\sqrt{\eta} \dot{{\bm x}})^\sschi_s  \cdot \frac{\partial }{\partial \bm{x}} + (\sqrt{\eta}  \dot{\bm p})^\sschi_s \cdot\frac{\partial }{\partial \bm{p}} 
+\hbar s\chi Q\frac{\bm \omega \cdot \bm E}{|\bm p|^2} -2 \hbar s\chi Q\frac{\bm p \cdot \bm\omega\, \bm p \cdot \bm E}{|\bm p|^4}\Big) f^s_{\sschi} (t,\bm x,\bm p)=0, \label{CKT3}
\end{equation}
where
\begin{eqnarray}
\sqrt{\eta}_{\, s}^{\, \sschi }   &=& 1 + \hbar s \chi \frac{\bm p\cdot \bm \omega}{|\bm p|^2} +\hbar \chi Q \frac{\bm B \cdot \bm p}{2|\bm p|^3},  \label{3e1} \\ 
(\sqrt{\eta}  \dot{\bm x})^\sschi_s &=&\bm v^\sschi_s + \hbar s\chi (\frac{1}{2}+\frac{k}{2}) \frac{\bm \omega}{|\bm p|}  +\hbar s\chi (\frac{1}{2}-\frac{k}{2})\frac{\bm p\cdot \bm \omega}{|\bm p|^3}\bm p \nonumber \\
&& + \hbar \chi Q\frac{\bm B}{2|\bm p|^2} + \hbar \chi Q\frac{\bm E \times \bm p}{2|\bm p|^3}   , \label{3e2}\\
(\sqrt{\eta} \dot{\bm p})^\sschi_s&=& sQ\bm E+ k |\bm p|\bm v_s^\sschi \times  \bm \omega  +s Q \bm v_s^\sschi \times  \bm  B \nonumber \\
& & - (k-1) \hbar s \chi  Q \frac{\bm p\cdot \bm \omega}{|\bm p|^3}\bm p \times \bm B   \nonumber \\
&& +\hbar s\chi Q^2\bm E \cdot \bm B \frac{\bm p}{2|\bm p|^3}  + \hbar Q\chi \frac{\bm p\cdot \bm \omega}{|\bm p|^2}\bm E .
\label{3e3}
\end{eqnarray}
We  ignore the ${\cal O}(\omega^2)$ terms.

The chiral
particle (antiparticle) number and current densities are defined by
\begin{eqnarray}
n^\chi_s& = &  \int \frac{d^3p}{(2\pi\hbar)^3} (\sqrt{\eta})_s^\sschi f^{eq,s}_{\sschi},\label{nil} \\
\bm j^\chi_s & = & \int \frac{d^3p}{(2\pi\hbar)^3}(\sqrt{\eta} \dot{\bm x})^\sschi_s f^{eq,s}_{\sschi}+
\bm \nabla \times  \int \frac{d^3p}{(2\pi\hbar)^3} \frac{s{\cal E}_s^\sschi  \bm p}{2 |\bm p|^3} f^{eq,s}_{\sschi} ,\label{jil}
\end{eqnarray}
where $f_{\sschi}^s \equiv f_{\sschi}^s (t,\bm x,\bm p).$

To accomplish the continuity equation satisfied by the 4-current density $j^{\sschi\mu}_s\equiv(n^\chi_s , \bm j^\sschi_s),$ let us calculate
\begin{eqnarray}
C&\equiv&\int d^3p\Big\{ \frac{\partial }{\partial t} \left[\sqrt{\eta}_{\, s}^{\, \sschi } f_{\sschi}^s\right]+  \frac{\partial }{\partial \bm{x}}\left[(\sqrt{\eta} \dot{{\bm x}})^\sschi_s f_{\sschi}^s\right] + \frac{\partial }{\partial \bm{p}} \left[(\sqrt{\eta}  \dot{\bm p})^\sschi_s f_{\sschi}^s\right] \Big\}\label{c} .
\end{eqnarray}
 Observe that (\ref{3e1}) and (\ref{3e2}) do not depend on time and spatial coordinates explicitly, so that  we have
\begin{eqnarray}
C &=&\int d^3p\Big\{ \Big( \sqrt{\eta}_{\, s}^{\, \sschi } \frac{\partial }{\partial t} + (\sqrt{\eta} \dot{{\bm x}})^\sschi_s  \cdot \frac{\partial }{\partial \bm{x}} + (\sqrt{\eta}  \dot{\bm p})^\sschi_s \cdot\frac{\partial }{\partial \bm{p}} \Big) f_{\sschi}^s+\frac{\partial (\sqrt{\eta}  \dot{\bm p})^\sschi_s}{\partial \bm{p}} f_{\sschi}^s\Big\}\label{cc} . \nonumber
\end{eqnarray}
 By employing the transport equation (\ref{CKT3}) one  gets
\begin{eqnarray}
C&=&\int d^3p \left[\frac{\partial (\sqrt{\eta}  \dot{\bm p})^\sschi_s}{\partial \bm{p}} 
-\hbar s\chi Q\Big(\frac{\bm \omega \cdot \bm E}{|\bm p|^2} -2 \frac{\bm p \cdot \bm\omega\, \bm p \cdot \bm E}{|\bm p|^4}\Big)\right] f_{\sschi}^s .\label{c1}
\end{eqnarray}
The derivative of (\ref{3e3}) leads to
\begin{eqnarray}
\int d^3p\ \frac{\partial (\sqrt{\eta}  \dot{\bm p})^\sschi _s}{\partial \bm{p}} f_{\sschi}^s&=&\hbar s\chi Q
\int d^3p \left(\frac{\bm \omega \cdot \bm E}{|\bm p|^2}-2\frac{\bm p \cdot \bm\omega\, \bm p \cdot \bm E}{|\bm p|^4} \right)
f_{\sschi}^s    \nonumber \\
&&+ 2\pi  \hbar Q^2\chi \bm E\cdot \bm B\int d^3p \delta (\bm p)  f_{\sschi}^s\label{pdot} .
\end{eqnarray}
By substituting the first term in (\ref{c1}) with (\ref{pdot}) one finally attains
\begin{eqnarray}
C&=& 2\pi\hbar Q^2\chi \bm E\cdot \bm B\int d^3p \delta (\bm p)  f^s_{\sschi} (t,\bm x,\bm p) .\label{pdott} 
\end{eqnarray}
On the other hand the last term of (\ref{c}) vanishes because it is a total derivative, thus   the continuity equation   is deduced:
\begin{equation} 
\label{ceqD0}
\frac{\partial n_s^\sschi}{\partial t} + \bm {\nabla} \cdot \bm j_s^\sschi= \frac{\chi Q^2}{(2\pi\hbar)^2}  \bm{{ E}}\cdot \bm{B} \ f^{eq,s}_{\sschi} (t,\bm x,\bm 0)  . 
\end{equation}

Let us introduce the vector and axial-vector currents
\begin{equation}
\bm j_{\ssV}=\bm j_\ssR +\bm j_\ssL, \ \ \  \bm j_{\ssA}=\bm j_\ssR -\bm j_\ssL, \label{rlc3}
\end{equation}
where $\bm j_{\ssR}=\sum_{s=\pm 1}\bm j^{1}_s $ and $\bm j_{\ssL}=\sum_{s=\pm 1}\bm j^{-1}_s .$
Let us choose the equilibrium distribution function as
\begin{equation}
f^{eq,s}_{\sschi} = \frac{1}{ e^{s({\cal E}_s^\sschi - \mu_\sschi )/ T } +1  } \cdot \label{norot}
\end{equation} 
Here $\mu_\chi =\mu + \chi \mu_5,$  where $\mu,\mu_5$  are the total and axial chemical potentials.  
By inspecting (\ref{3e2}) one observes that the currents  are linear in the magnetic field and vorticity:
\begin{eqnarray}
\bm j_\ssV &= &  \xi_\ssB \bm B+ \xi \bm \omega ,\label{3DjV}\\
\bm j_\ssA &= &  \xi_{\ssB 5} \bm B+ \xi_5 \bm \omega  . \label{3DjA}
\end{eqnarray}
The coefficients of the magnetic field are calculated as
\begin{equation}
\xi_{{\ssB}} =\frac{Q\mu_{\ssbes} }{2  \pi^2 \hbar^2} , \ \ 
\xi_{\ssBb} =  \frac{Q\mu}{2 \pi^2 \hbar^2}  .\nonumber
\end{equation}
Thus the magnetic terms in (\ref{3DjV}) and (\ref{3DjA}), respectively, generate the chiral magnetic and chiral separation effects. The vorticity terms in (\ref{3DjV}) and (\ref{3DjA}), respectively, generate  the chiral vortical  and local polarization effects  correctly for
\begin{equation}
\xi = \frac{\mu  \mu_{\ssbes}}{\pi^2\hbar^2} , \ \  \xi_{\ssbes} = \frac{T^2}{6 \hbar^2} + \frac{\mu^2+{\mu_{\ssbes}}^2}{2  \pi^2 \hbar^2}  .\label{omco}
\end{equation}
However, the  coefficients $\xi$ and $\xi_5$ depend on $k.$ One can show that (\ref{omco}) results 
as far as the condition 
\begin{equation}
\frac{2}{3}+\frac{k}{3}=1,
\label{k1}
\end{equation}
is satisfied.  This yields  $k=1.$ This value of $k$ is in accord with the formalism  considered in \cite{dk}.  

We do not deal with the equilibrium distribution function for a rotating fluid because in constructing the  CKE (\ref{cke1})  we have not considered  rotation of the reference  frame.  However, in the absence of modification terms one should work with the equilibrium distribution function of a rotating fluid in the comoving frame of the fluid  \cite{soncollisionckt,hpy2}
\begin{equation}
\label{frc}
f^{eq,s}_{\sschi} =\frac{1}{e^{s\left(u\cdot p -\mu_\chi +\frac{\hbar }{2}S^{\mu\nu}\partial_\mu u_\nu\right)/T} +1 },
\end{equation}
where $S^{\mu\nu}$ is given in (\ref{smunu}). In the frame  $u^\alpha=(1,\bm 0)$ and  $\omega^\alpha =(0, \bm \omega),$  the distribution function (\ref{frc}) yields 
\begin{equation}
f^{eq,s}_{\sschi} = \frac{1}{ e^{s({\cal E}_s^\sschi - \mu_\sschi -\hbar  \sschi   \hat{\bm p} \cdot \bm \omega /2)/ T } +1  } .
\end{equation}
If one adopts this equilibrium distribution function
the values in (\ref{omco})   are  acquired when  the condition 
\begin{equation}
\frac{5}{6}+\frac{k}{3}=1,
\label{k2}
\end{equation}
is fulfilled, which yields $k=1/2.$ Obviously, for  the original CKE  where $k=0,$  the condition (\ref{k2}) cannot be satisfied. Therefore, we conclude that without the modification the 3D formalism obtained from (\ref{cke1}) does not generate the CVE correctly. This result is pertinent to the transport equation  in Minkowski space with the condition $n_\mu =u_\mu,$ where $u_\mu$ and $\omega_\mu$ are given as in (\ref{cuo}). Obviously, one can choose either $n_\mu$ or $u_\mu$ in a different manner, e.g.  $u_\mu=(1, \bm{x}\times \bm{\omega}).$ For each choice one should derive the resulting 3D transport equations by integrating (\ref{CKTn}) over $p_0.$

As it  was mentioned in Introduction there also exists a curved spacetime formulation of chiral kinetic equation  \cite{lgmh}  where the Coriolis force and CVE are generated correctly without  a  need for  modification.  However, this formalism leads to a 3D  kinetic theory  which depends on $\bm x$  explicitly  \cite{dk2019}, in contrary to the 3D kinetic theories obtained here (\ref{3e1})-(\ref{3e3}) or as it has been  derived in \cite{dk}.

\section{Kinetic Equations of Massive Fermions}
\label{sec-mass}
The  equations  (\ref{real1})-(\ref{imag5}) which should be satisfied by the components of Wigner function in Clifford algebra basis  are reducible: The field equations (\ref{real1}), (\ref{real4}) and (\ref{real5}) can be employed to express the fields $\mathcal{F},\, \mathcal{P},\, \mathcal{S}_{\mu \nu}$ in terms of the vector and axial-vector fields, $\mathcal{V}_{\mu},\, \mathcal{A}_{\mu}.$ Following the procedure given in \cite{oh} and \cite{hhy} the rest of field equations (\ref{real1})-(\ref{imag5}), can be shown to yield  
\begin{eqnarray}
 \tilde{\nabla} \cdot \mathcal{V}&=&0, \label{mathcalV}\\ 
 {p\cdot \mathcal{A} }&=&0,  \label{I} \\
	\left(p^{2}-m^{2}\right) \mathcal{V}_{\mu} &=&-\hbar (Q\tilde{F}_{\mu \nu}+\zeta \tilde{w}_{\mu \nu}) \mathcal{A}^{\nu}, \label{II} \\
p_{\nu} \mathcal{V}_{\mu}-p_{\mu} \mathcal{V}_{\nu}&=&6\frac{\hbar}{2} \epsilon_{\mu \nu \rho \sigma}  \tilde{\nabla}^{\rho} \mathcal{A}^{\sigma} ,\label{III}\\
\left(p^{2}-m^{2}\right) \mathcal{A}^{\mu}&=&\frac{\hbar}{2} \epsilon^{\mu \nu \rho \sigma} p_{\sigma}  \tilde{\nabla}_{\nu} \mathcal{V}_{\rho}, \label{IV} \\
p \cdot  \tilde{\nabla} \mathcal{A}^{\mu}+(QF^{\nu \mu}+\zeta w^{\nu \mu}) \mathcal{A}_{\nu}&=&\frac{\hbar}{2} \epsilon^{\mu \nu \rho \sigma}\left(Q\partial_{\sigma} F_{\beta \nu}+\partial_{\sigma} \zeta w_{\beta \nu}\right) \partial_{p}^{\beta} \mathcal{V}_{\rho} \label{mathcalA},
\end{eqnarray}
where at most ${\mathcal O}(\hbar)$ terms are taken into account.  
We can derive the
semiclassical kinetic equations   resulting from (\ref{mathcalV})-(\ref{mathcalA})  by substituting $QF_{\mu \nu}$ with $QF_{\mu \nu}+\zeta w_{\mu \nu}$  in the formalism which has been given   in \cite{hhy} for $\zeta =0.$  
 First one solves (\ref{II}), (\ref{III}) for ${\mathcal V}_\mu$  up to $\hbar$-order, then uses it in (\ref{mathcalV}) and gets
the kinetic equation of the vector field:
\begin{eqnarray}
	\begin{aligned}
		& \delta\left(p^{2}-m^{2}\right)  p \cdot \tN f_{V}+ \hbar  \delta\left(p^{2}-m^{2}\right) \Biggl\{ \Big(\frac{n^\alpha}{p \cdot n} S_{a(n)}^{\mu \nu} (Q F_{\mu \alpha}+\zeta w_{\mu \alpha}) + \partial_{\mu} S_{a(n)}^{\mu \nu} \Big) \tN_{\nu}  \\
	&+S_{a(n)}^{\mu \nu}\partial_{\mu} (QF_{\rho \nu} +\zeta w_{\rho \nu})  \partial_{p}^{\rho} +\frac{ \epsilon^{\mu \nu \alpha \beta}}{2} \Biggr[  \tN_{\mu}\left(\frac{n_{\beta}}{p \cdot n}\right)\left[\tN_{\nu} a_{\alpha}+Q F_{\nu \alpha}+ \zeta w_{\nu \alpha}\right]\\
		&+\frac{n_{\beta}}{p \cdot n}\left( \partial_{\mu} \left( QF_{\rho \nu}+ \zeta  w_{\rho \nu} \right)\partial_{p}^{\rho} a_{\alpha}+\left[\tN_{\nu} a_{\alpha}-(Q F_{\rho \nu} + \zeta w_{\rho \nu})\partial_{p}^{\rho} a_{\alpha}\right] \tN_{\mu}\right) \Biggr]\Biggr\}  f_{A}\\
			&-\hbar \frac{\delta^{\prime}\left(p^{2}-m^{2}\right)}{2 (p \cdot n)} \epsilon_{\mu \nu \alpha \rho} n^\nu (Q F^{\alpha \rho}+ \zeta w^{\alpha \rho})  [p \cdot \tN ({a}^{\mu} f_{A})+F^{\sigma \mu} {a}_{\sigma} f_{A}] =0.
	\end{aligned}
	\label{SKE}
\end{eqnarray}
Here $f_V,f_A$ are vector and axial distribution functions.
\begin{equation}
S^{\mu \nu}_{\ssa \ssbfn}=
\frac{1}{ 2 n \cdot p}  \epsilon^{\mu \nu \rho \sigma} a_\rho  n_\sigma
\label{S_a}
\end{equation}
is the spin tensor and  $a_\mu$ is the spin four-vector which is defined to satisfy the constraint
\begin{equation}
\label{adq}
a\cdot p =p^2-m^2.
\end{equation}
To derive  the other kinetic equations  which are needed to determine the dynamical degrees of freedom  $f_V, f_A, a_\mu,$ one solves (\ref{I}) and (\ref{IV}) for ${\mathcal A}^\mu$ up to ${\mathcal O}(\hbar)$ with the help of Wigner function of the quantized free Dirac fields \cite{hhy}. Plugging this solution  into (\ref{mathcalA})  leads to
\begin{eqnarray}
	\begin{aligned}
		& \delta\left(p^{2}-m^{2}\right) \left(p \cdot \tN \left(a^{\mu} f_{A}\right)+(Q F^{\nu \mu}+\zeta w^{\nu \mu}) a_{\nu} f_{A}\right)\\ 
		&+\hbar  \delta\left(p^{2}-m^{2}\right)  \Biggl\{ p^{\mu} S_{m(n)}^{\rho \nu}\partial_{\rho} \left(F_{\beta \nu} + \zeta w_{\beta \nu}\right) \partial_{p}^{\beta}+p^{\mu} \Big(\partial_{\alpha} S_{m(n)}^{\alpha \nu} +\frac{S_{m(n)}^{\alpha \nu} (Q F_{\alpha \beta}+\zeta w_{\alpha \beta}) n^\beta}{p \cdot n+m} \Big)\tN_{\nu} \\
		&+ \frac{\epsilon^{\mu \nu \alpha \beta}}{2(p \cdot n+m)} \Big[\frac{m^2 n_{\beta}+mp_{\beta}}{p \cdot n+m} \left[(Q F_{\alpha \sigma}+\zeta w_ {\alpha \sigma})n^\sigma - \partial_{\alpha }(p \cdot n)\right]  + m^2 \left(\partial_{\alpha} n_{\beta}\right) \Big]\tN_{\nu} \Biggr\}f_{V}\\
		&- \hbar \frac{\delta^{\prime}\left(p^{2}-m^{2}\right) }{(p \cdot n+m)} \Big(p^{\mu} p_\alpha n_\beta (Q \tilde{F}^{\alpha \beta}+\zeta \tilde{w}^{\alpha \beta})-\left(m^2 n_{\beta}+mp_{\beta}\right) (Q\tilde{F}^{\mu \beta}+\zeta \tilde{w}^{\mu \beta})\Big) p \cdot \tN f_{V} =0,
	\end{aligned}
	\label{AKE}
\end{eqnarray}
where 
\begin{equation}
S^{\mu \nu}_{\ssm \ssbfn}=
\frac{1}{ 2( n \cdot p +m)}  \epsilon^{\mu \nu \rho \sigma} p_\rho  n_\sigma .
\label{S_p}
\end{equation}
Once we get these kinetic equations, we impose the equations of motion  resulting from (\ref{SEM}) and (\ref{sF}). Thus, from now on we deal with $F_{\mu \nu}$ expressed in terms of the electromagnetic fields $E_\mu, B_\mu$ as in (\ref{fEM}) and  the field strength $  w^{\mu\nu}$ expressed in terms of the vorticity $\omega_\mu$ and the { energy per particle $u\cdot p,$ }as in (\ref{wab1}). To keep the discussion general let us introduce
\begin{equation}
\zeta w^{\mu\nu}  \equiv  \varLambda w^{\mu\nu}_\ssC+2\varepsilon   u\cdot p\,\Omega^{\mu\nu} . 
\label{zetaw}
\end{equation}
How to fix the constants $\varepsilon$ and $\varLambda$ will be discussed  later.
By making use of (\ref{Ommunu}) and the Schouten identity,  (\ref{zetaw}) can be written equivalently as
\begin{equation}
\zeta w^{\mu\nu}=	\epsilon^{\mu \nu \sigma \rho} \omega_\rho (u\cdot p\, u_\sigma (2 \varepsilon+\varLambda) - \varLambda p_\sigma).
\label{zetaw-son}
\end{equation}	
The kinetic equations (\ref{SKE}) and (\ref{AKE}) form the 4D collisionless kinetic theory of massive fermions in the presence electromagnetic fields and vorticity. 
We  would like to derive the nonrelativistic  kinetic theory arising from this covariant formulation.  For this purpose let us  extract the scalar kinetic equation stemming from (\ref{AKE}) by projecting it on $n_\mu:$
\begin{eqnarray}
	&&\delta (p^2-m^2)n_\mu\left[ p\cdot \tN (a^\mu f_A) -(QF^{\mu \sigma}+\zeta w^{\mu \sigma})a_\sigma f_A\right] +\hbar (p\cdot n +m) \delta (p^2-m^2)\Bigg[(\partial_{\sigma }S^{\sigma\nu}_{\ssm \ssbfn}) \tN_\nu \nonumber\\
	&&+\frac{n^\sigma }{p\cdot n +m}(QF_{\mu \sigma}+\zeta w_{\mu \sigma})S^{\mu \nu}_{\ssm \ssbfn} \tN_\nu + S^{\mu \nu}_{\ssm \ssbfn} [\partial_\mu (QF_{\alpha \nu}+\zeta w_{\alpha \nu})]\partial_p^\alpha \Bigg] f_V \label{SAKE} \\
	&& +\hbar m\frac{\delta (p^2-m^2)}{ 2( n \cdot p +m)}\epsilon^{\mu \nu \alpha \rho}(mn_\mu +p_\mu)(\partial_\alpha n_\rho)\tN_\nu f_V -\hbar \delta^\prime (p^2-m^2) p_\alpha n_\beta  (Q\tilde{F}^{\alpha \beta}+\zeta \tilde{w}^{\alpha \beta})p \cdot \tN f_V =0. \nonumber
\end{eqnarray}
We will show that  (\ref{SKE}) and (\ref{SAKE}) can be combined to derive   scalar kinetic equations which generate correct dispersion relations of Dirac fermions coupled to electromagnetic fields in a  frame rotating with the angular velocity $\bm \omega .$

\subsection{3D Massive kinetic theory}

To establish the  3D kinetic theory  arising from the relativistic kinetic equations (\ref{SKE}) and (\ref{SAKE}), one should determine  the form of  $a_\mu ,$ consistent with the kinetic equations and the constraint (\ref{adq}). However we are only interested in attaining small mass corrections to the chiral effects. Hence, let us deal with $a_\mu$ 
derived  from the Wigner function of the free Dirac fields \cite{hhy}. In this case it is defined to satisfy
\begin{eqnarray}
&a\cdot n f_A=(p\cdot n +m)f_A,& \ \ \ a_{\perp \mu} f_A=p_{\perp \mu}f_A-m S_\mu,
\end{eqnarray}
where $S_\mu$ is the spin four-vector  of the Dirac wave-function and $a_{\perp}\cdot n= p_{\perp}\cdot n=0 .$
Thus, it can be expressed as
\begin{equation}
a_\mu=mn_\mu+p_\mu -mS_\mu \nonumber .
\end{equation}
$
S\cdot p =p\cdot n +m,
$
so that  (\ref{adq}) is fulfilled. 
Moreover, let  $S_\mu$ be given as in the massless case:
$$
S^\mu= -\frac{p_{\perp \mu}}{p\cdot n} f_A.
$$
Therefore, we set
\begin{equation}
\label{amu}
a_\mu =(1+\frac{m}{p\cdot n}) p_\mu.
\end{equation}
Because of dealing with $a_\mu$ at the zeroth-order in electromagnetic fields and vorticity, we need to keep the terms which are  at most linear in $F_{\mu\nu}$ and $w_{\mu\nu}$  in the kinetic equations   (\ref{SKE}) and (\ref{SAKE}). For simplicity, we consider the fields satisfying $\partial_{\mu}F_{\nu \sigma}=0$ and $\partial_\mu \omega_\nu =0.$

To establish the 3D kinetic theory  by integrating over $p_0,$ we have to work in  the comoving frame by setting
\begin{equation}
n_\mu=u_\mu.
\label{cmf}
\end{equation}
Let us accomplish the kinetic equations following from (\ref{SKE}) and (\ref{SAKE}) in the comoving frame (\ref{cmf}), by keeping at most the terms linear in $E_{\mu},$ $B_{\mu}$ and $\omega_\mu.$ 
The kinetic equation of the vector field (\ref{SKE}) reads
	\begin{eqnarray}
		&&\delta (p^2-m^2) p\cdot \tN f_V \nonumber\\
		&&+\hbar \delta (p^2-m^2)\Bigg[ (\partial_{\sigma }S^{\sigma\nu}_{\ssa \ssbfu})  +\frac{ S^{\mu \nu}_{\ssa \ssbfu} }{p\cdot u}(Q E_{\mu}+\varLambda \, \Omega_{\mu \alpha} p^\alpha)  -\hbar m \ \frac{\epsilon^{\nu \mu \alpha \beta}}{ 2( u \cdot p)^3}u_\beta   p_\alpha (\partial_\mu u_\sigma) p^\sigma   
		\Bigg]\partial_\nu f_A \nonumber\\
		&&	- \frac{\delta^\prime (p^2-m^2)}{ 2( u \cdot p )^2} (u \cdot p +m)   \big[Q B \cdot p+ 2 \alpha (u\cdot p) (\omega \cdot p)\big]p\cdot \partial f_A =0,
		\label{SKE-son}
\end{eqnarray} 
with the spin tensor
\begin{equation}
S^{\mu \nu}_{\ssa \ssbfu}= \frac{1}{ 2 u \cdot p }  \epsilon^{\mu \nu \rho \sigma} p_\rho  u_\sigma
+\frac{m}{ 2 (u \cdot p)^2 }  \epsilon^{\mu \nu \rho \sigma} p_\rho  u_\sigma. \nonumber
\end{equation}
The left-hand side of (\ref{SAKE})  can be written with an overall   $ (u\cdot p +m) $ factor which is obviously nonvanishing. Hence (\ref{SAKE})  yields the kinetic equation 
	\begin{eqnarray}
		&&\delta (p^2-m^2) \Bigg[p\cdot \tN f_A -\frac{m}{p\cdot u (u\cdot p +m)} \Big( (\partial_{\alpha }u_\beta) p^\alpha p^\beta- p_\mu u_\nu QF^{\mu \nu} \Big) f_A \Bigg]  \nonumber\\
		&&+\hbar \delta (p^2-m^2) \Bigg[\partial_{\sigma }S^{\sigma\nu}_{\ssm \ssbfu} +\frac{ S^{\mu \nu}_{\ssm \ssbfu} }{p\cdot u +m}(Q E_{\mu}+\varLambda \, \Omega_{\mu \alpha} p^\alpha) +\frac{\epsilon^{\mu \nu \alpha \rho}(\partial_\alpha u_\rho) }{ 2( u \cdot p +m)^2}(m^2 u_\mu +m p_\mu)
		\Bigg] \partial_\nu f_V \nonumber\\
		&& 	- \frac{\delta^\prime (p^2-m^2)}{ 2( u \cdot p +m)}\big[Q B \cdot p+ 2 \alpha (u\cdot p) (\omega \cdot p)\big] p\cdot \partial f_V=0. \label{SAKE-son1}
	\end{eqnarray}
By expanding  the denominators  for small mass with  $m/u\cdot p\ll1,$ and ignoring the $m^2/(u\cdot p)^2$ and higher order terms,  (\ref{SAKE-son1}) can be written as
\textbf{\begin{eqnarray}
	&&\delta (p^2-m^2) \Bigg[p\cdot \tN f_A +\frac{m}{(u\cdot p )^2} QE \cdot p f_A \Bigg]  \nonumber\\
	&&+\hbar \delta (p^2-m^2) \Bigg[ \partial_{\sigma }S^{\sigma\nu}_{\ssm \ssbfu} +\left(1-\frac{m}{u\cdot p}\right)\frac{S^{\mu \nu}_{\ssm \ssbfu} }{p\cdot u }(Q E_{\mu}+\varLambda \,\Omega_{\mu \alpha} p^\alpha)+\frac{ m p_\mu}{ ( u \cdot p)^2} \tilde{\Omega}^{\mu \nu}
	\Bigg] \partial_\nu f_V \nonumber\\
	&& 	- \hbar \left(1-\frac{m}{u\cdot p}\right)\frac{\delta^\prime (p^2-m^2)}{  u \cdot p } \big[Q B \cdot p+ 2 \varepsilon (u\cdot p) (\omega \cdot p)\big] p\cdot \partial f_V=0 .\label{SAKE-son}
	\end{eqnarray}}
In the small mass limit we have
\begin{equation}
S^{\mu \nu}_{\ssm \ssbfu} \approx  \frac{1}{ 2 u \cdot p }  \epsilon^{\mu \nu \rho \sigma} p_\rho  u_\sigma
-\frac{m}{ 2 (u \cdot p)^2 }  \epsilon^{\mu \nu \rho \sigma} p_\rho  u_\sigma.  \nonumber
\end{equation}

Then, by summing and subtracting  (\ref{SKE-son}) and  (\ref{SAKE-son})  we acquire the kinetic equations 
\begin{eqnarray}
	&&\delta \left( p^2-m^2 -\chi \hbar   \frac{Qp\cdot B}{u\cdot p} -2 \varepsilon \chi  \hbar p\cdot \omega\right)  \Biggl\{ p\cdot \tN \nonumber\\
	&&+\chi \frac{\hbar}{2(u \cdot p)^2}  \epsilon^{\mu \nu \alpha \beta} p_\alpha \Big[Qu_\beta E_\mu  +  u \cdot p\, \Omega_{\mu \beta}+(\varLambda-1) u_\beta \Omega_{\mu \sigma} p^\alpha  \Big]  \partial_\nu 
	\label{eom-fR} \\
	&&- \chi \frac{\hbar}{4 (u \cdot p)^3} \epsilon^{\mu \nu \alpha \beta} m \Big[ p_\alpha u_\beta ( Q E_\mu + \varLambda \ \Omega_{\mu \sigma} p^\sigma)- u \cdot p\,  p_\mu \Omega_{\alpha \beta} \Big] \partial_\nu \nonumber\\
	&&- \frac{\hbar}{4(u \cdot p)^3}  \epsilon^{\mu \nu \alpha \beta}  m \ p_\alpha u_\beta \Omega_{\mu \sigma} p^\sigma\partial_\nu 
	-  \frac{m}{2(u \cdot p)^2} Q E\cdot p   \Biggr\} f_{\chi} =C_{(-\chi)}, \nonumber
\end{eqnarray}
where
\begin{eqnarray}
C_{\chi}=&&- \chi \frac{\hbar \delta \left( p^2-m^2 \right)}{2(u \cdot p)^2}  \epsilon^{\mu \nu \alpha \beta} \Biggr\{ \ m \ p_\alpha \Big[\Omega_{\mu \beta} -2\,\Omega_{\mu \sigma} p^\sigma u_\beta  \Big] \partial_\nu  \nonumber\\
&&- \frac{1}{4(u \cdot p)^3}  m \Big[  p_\alpha u_\beta (Q E_\mu + \varLambda \, \Omega_{\mu \sigma} p^\sigma)- u \cdot p\, p_\mu\Omega_{\alpha \beta} \Big] \partial_\nu \nonumber\\
&&+ \frac{m }{2(u \cdot p)^2}  Q E \cdot p  \Biggr\}  f_{\chi} \nonumber\\
&& - m \delta^\prime (p^2-m^2)  ( Q p\cdot B + 2\varepsilon (u\cdot p) (p \cdot \omega)) p \cdot \partial f_{\chi}  .
\label{C_R/L}
\end{eqnarray}
Here  $f_\chi \equiv f_{R/L}=(f_V+\chi f_A)/2$ are the right-handed and left-handed distribution functions. The right-hand side of (\ref{eom-fR}) vanishes for both  $m=0$ and  $\hbar=0,$ but its  left-hand side  is non-zero whether $\hbar =0$ or  $m=0.$ 
Moreover, the distribution functions appearing on  the left- and the right-hand sides possess opposite chirality.  Hence we can consider (\ref{eom-fR}) as the transport equation of $f_\chi$ which shows up  on the left-hand side  where its right-hand side is due to presence of the opposite chirality distribution function $f_{-\chi},$ for massive fermions. 

To derive the nonrelativistic kinetic equations by integrating  (\ref{eom-fR}) over $p_0,$ we  consider distribution functions composed as in (\ref{f_fd}) and choose the  frame  $u_\mu=(1, \bm 0),$ $\omega^\mu=(0, \bm \omega).$  In this frame the Dirac delta function yields  the  dispersion relations
\begin{equation}
\label{edr}
	p_{0s}^{\chi}=sE_p\left[1-s\chi \hbar \ \bm b \cdot \big(Q\bm B+2 \varepsilon E_p \bm \omega\big)\right],
\end{equation}
where $E_p=\sqrt{\boldsymbol{p}^2-m^2}$ and  \mbox{$\bm b \equiv \bm p/2 E_p^3.$ }
 Calculation of the $p_0$ integral yields
\begin{equation}
\big( \sqrt{\eta}_{\, \chi}^{\, s} \frac{\partial }{\partial t} + (\sqrt{\eta} \dot{{\bm x}})_{\chi}^{s}  \cdot \frac{\partial }{\partial \bm{x}} + (\sqrt{\eta}  \dot{\bm p})_{\chi}^{s} \cdot\frac{\partial }{\partial \bm{p}}\big) f_{\chi}^{s} (t,\bm x,\bm p)  + \frac{m }{2 E_p^3} sQ \bm E \cdot \bm p)  f_{\chi}^s (t,\bm x,\bm p)  =C_{(-\chi)},
\label{3d-keR} 
\end{equation}
where
\begin{eqnarray}
	\sqrt{\eta}_{\, \chi}^{\, s}&=&  1+ \hbar \chi \bm b \cdot (Q\bm B +  s(2E_p+m)  \bm \omega), \\ 
	(\sqrt{\eta} \dot{{\bm x}})_{\chi}^{s}  
	&=& \frac{\bm p}{E_p} \Big\{1+2 \hbar \chi Q\bm b \cdot \bm B +  \frac{\hbar s}{2} [2 \chi E_p (2 \varepsilon +(1-\varLambda) )+( \chi +1)m]  \ \bm b \cdot \bm \omega \Big\} \nonumber \\
	&& +\hbar \chi Q \Big(1-\frac{m}{E_p}\Big) \bm E \times \bm b+\hbar \chi s \Big(1-\frac{m}{2 E_p}\Big) \frac{\bm \omega}{E_p} \nonumber \\
	&&-\hbar \bm \omega (\bm b \cdot \bm p) \Big( \frac{ s m}{2 E_p} ( \chi \varLambda +1) + \chi (1-\varLambda) \Big),  \label{xdot-m}\\
	(\sqrt{\eta}  \dot{\bm p})_{\chi}^{s}&=& s Q\bm E+ \frac{\bm p}{E_p}  \times \left[eQ \bm B +(2\varepsilon+\varLambda)E_p \bm \omega\right]. \label{pdm}
\end{eqnarray}
The right-hand side of (\ref{3d-keR}) which vanishes for $m=0$ can be considered  as the correction terms  due to the presence of opposite chirality  fermions in the massive case.  Terms which vanish for $m=0$ can also be interpreted as collision terms due to interaction of electromagnetic fields and vorticity with  the spin of massive fermions \cite{Wang_2021}.

Let us fix the values of $\varepsilon$ and $\varLambda .$ Comparing the  dispersion relations (\ref{edr}) with the Hamiltonian of Dirac particles coupled to the magnetic field in rotating coordinates   in  the helicity basis \cite{dky,dk-m},  one observes that  $\varepsilon=1/2.$ Similar energy   relations were also obtained for chiral particles in \cite{cssyy,lgmh}. Now, by inspecting (\ref{pdm}) we see that $\varLambda=1$ for reproducing the Coriolis force correctly. This choice is also consistent with the massless case considered in the  section \ref{sec-cke}.

Inserting   the spatial velocity   (\ref{xdot-m})  into the following definition 
\begin{equation} 
 \bm j^\chi_s =  \int \frac{d^3p}{(2\pi\hbar)^3}(\sqrt{\eta} \dot{\bm x})^\sschi_s f^{eq,s}_{\sschi},
\label{j-1}
\end{equation}
where $f^{eq,s}_{\sschi}$ is given in  (\ref{norot}) by substituting  ${\cal E}_s^\sschi $ with (\ref{edr}), we write the particle number axial-vector and vector current densities  as 
\begin{equation}
\bm j _\ssA =\sum_{s\chi}\chi \bm j^\chi_s,\ \ \ \bm j _\ssV =\sum_{s\chi} \bm j^\chi_s .
\end{equation}
They  can be decomposed as follows
\begin{equation} 
\boldsymbol{j}_{\ssA, \ssV}^{B, \omega}(\boldsymbol{x}, t)=\overline{\sigma}_{\ssA, \ssV}^{B} \boldsymbol{B}+\overline{\sigma}_{\ssA, \ssV}^{\omega} \boldsymbol{\omega},
\end{equation}
with
\begin{eqnarray}
	\begin{aligned}
	\bar{\sigma}_{\ssA, \ssV}^{B}=&\frac{Q}{6 \pi^{2} \hbar^{2}} \int d|\bm{p}|\left\{\frac{|\bm{p}|^{4}}{E_p^{4}} f_{\ssA, \ssV}^{0}-\frac{|\boldsymbol{p}|^{4}}{2 E_p^{3}} \frac{\partial f_{\ssA, \ssV}^{0}}{\partial E_p}\right\} \\
	\bar{ \sigma}_{\ssA, \ssV}^{\omega}=&\frac{1}{2 \pi^{2} \hbar^{2}} \int d|\bm{p}|  \Bigg( \Big(\frac{|\bm{p}|^{2}}{E_p}+\frac{|\bm{p}|^{4} (\varepsilon+\varLambda-1)}{3 E_p^3} -\frac{ (3\varLambda-1)|\bm{p}|^{4} m}{12 E_p^4}-\frac{m |\bm{p}|^{2}}{2E_p^2} \Big) f_{\ssA, \ssV}^{0}\\
	&- \frac{|\bm{p}|^{4} m}{6 E_p^4} f_{V,A}^{FD}-\frac{\varepsilon|\bm{p}|^{4}}{3 E_p^2} \frac{\partial f_{\ssA, \ssV}^{0}}{\partial E_p}\Bigg), \label{sigma_Bw}
\end{aligned}
\end{eqnarray}
where $f_{\ssA, \ssV}^{0}$ are defined by 
\begin{equation}
	f_{\ssA, \ssV}^{0}=\sum_{s}\left\{\frac{1}{e^{s\left[E_p-\mu_{R}\right] / T}+1} \pm \frac{1}{e^{s\left[E_p-\mu_{L}\right] / T}+1}\right\}.
\end{equation}

For the sake of simplicity let us deal with  $\mu_{\ssR}=\mu_{\ssL}=\mu$. While $\bar{\sigma}_{\ssV}^{\ssB}$ vanishes under this condition, $\bar{ \sigma}_{\ssV}^{\omega}$ is 
\begin{equation}
	\bar{ \sigma}_{\ssV}^{\omega} \underset{T = 0}{=}-\frac{m}{6\pi^2\hbar^2}\sqrt{\mu^2-m^2}\left(1+\frac{m^2}{2 \mu^2}\right)\theta(\mu-m),
\end{equation}
at zero temperature. By performing the integrals at zero temperature, the coefficients of   axial-vector current density are calculated as
\begin{eqnarray}
	\begin{aligned}
		\bar{\sigma}_{\ssA}^{B} \underset{T = 0}{=} &\ \frac{1}{2 \pi^{2} \hbar^{2}} \sqrt{\mu^2-m^2} \theta(\mu-m) , \\
		\bar{\sigma}_{A}^{\omega}  \underset{T= 0}{=} &\ \frac{1}{  \pi^{2} \hbar^{2}} \Big[ \mu  \sqrt{\mu^2-m^2} \Big(\frac{2+3\varepsilon+\varLambda}{6}-(3\varLambda-1)\frac{m^3}{24 \mu^3} \\
		&-(3\varLambda+5)\frac{m}{12 \mu} +(\varepsilon+\varLambda-1) \frac{m^2}{3 \mu^2} \Big) \\
		&-m^2 \ln \left(\mu/m + \sqrt{\mu^2/ m^2-1}\right) \Big]\theta(\mu-m).
	\end{aligned}
\end{eqnarray}
$ 	\bar{\sigma}_{A}^{B}$    is in accord with the one derived by means of Kubo formula \cite{gmsw,bgb1,lykubo}. For $\varepsilon=1/2, \varLambda=1$ we have
\begin{eqnarray}
\begin{aligned}
\bar{\sigma}_{A}^{\omega}  \underset{T= 0}{=} &\ \frac{1}{  \pi^{2} \hbar^{2}} \Big[ \mu  \sqrt{\mu^2-m^2} \Big(\frac{3}{4}-\frac{m^3}{12 \mu^3} -\frac{2m}{3 \mu} +\frac{m^2}{6 \mu^2} \Big) \\
&-m^2 \ln \left(\mu/m + \sqrt{\mu^2/ m^2-1}\right) \Big]\theta(\mu-m).
\end{aligned}
\end{eqnarray}

Note that when the massless limit is considered one should also set $\varepsilon=0.$ 
For instance  for small  $m,$ by setting $\varepsilon=0,$ one obtains
\begin{equation}
\bar{\sigma}_{\ssA} \approx \frac{1}{ 2 \pi^{2} \hbar^{2}} \left(\mu^2 -\frac{m^2}{2}\right).
\end{equation}
This is  the correction obtained in \cite{FlFu,lykubo}  in the $\mu=0$ limit. 
 
As we have already mentioned there is not a unique method of defining  kinetic transport  equations  for massive fermions   beginning with (\ref{real1})-(\ref{imag5}).  Hence there is no consensus in obtaining the mass corrections to the chiral magnetic and vortical effects. Here we established the massive  kinetic equation from the covariant formulation which directly generates the chiral transport equation when one sets  $a^\mu=p^\mu$ and $m=0.$ In fact we have chosen the spin four-vector as in (\ref{amu}) which is consistent with  the massless limit. Therefore the corrections to the chiral effects are apparent in contrary to the other formalisms which are suitable to discuss large mass limit \cite{wsswr,dk}.

The massless limit can also be discussed starting from (\ref{SKE}) and (\ref{SAKE}). For $m=0$ we need to set $\varepsilon =0$ in   (\ref{zetaw}) and fix  the spin four-vector  as $a_\mu=p_\mu.$ One can then observe that  (\ref{SAKE}) yields a scalar kinetic equation up to an overall $p^\mu$ factor. A similar scalar kinetic equation follows from  (\ref{SKE}). By adding and subtracting these scalar kinetic equations one obtains  the chiral kinetic equation (\ref{cke1}) for $n_\mu=u_\mu$ and ${\varLambda} =k.$ Therefore the 3D chiral theory coincides with the one obtained in section 5,  in particular particle number current density satisfies the anomalous divergence as in (\ref{ceqD0}).

\section{Conclusions}
\label{sec-conc}

We demonstrated that the scalar field $\phi$ and the vector-field $\eta_\mu$ represent the fermionic fluid  in the presence of  the Coriolis force due to the nonvanishing vorticity.  This is achieved by showing that the equations of motion acquired by variation of the action (\ref{sF})   with respect to   $\phi$ and $\eta_\alpha$  are equivalent to relativistic Euler equations of  a fluid with the Coriolis force.  Moreover, this formalism provided the field strength of the vector field $\eta_\alpha$ in terms of the  { specific} enthalpy and   fluid vorticity.  

Then we considered  the action of  Dirac  spinors  coupled  to the vector fields $A_\mu$ and $\eta_\mu.$ By virtue of its gauge invariances we derived the QKE satisfied by the Wigner function, (\ref{qke}),  by generalizing the formalism of  \cite{vge}.   In fact, one of the main  results accomplished in this work is to show that when one deals with the fluids  the original QKE \cite{vge} should  be modified  adequately. This modification has already been  introduced  in \cite{dk,dk-m} in an ad hoc manner. Here we obtained it in a systematic way from an underlying  action.

The Clifford algebra generators are employed to  decompose the Wigner function in terms of  field components and the semiclassical equations  of the component fields which follow from QKE are presented. Then to derive KTE one can  proceed in two different ways depending on when to impose the  on-shell conditions of  fields representing the fluid.  In \cite{dk,dk-m}  we have derived the semiclassical kinetic equations by imposing the on-shell conditions from the start.  In contrary here we first acquired the semiclassical transport equations and then let the fields $\eta_\alpha,\, \phi$ be on-shell, so that   $w_{\mu \nu}$  is expressed in terms of the vorticity and fluid 4-velocity.  This approach furnishes novel kinetic transport equations for either  massless or massive fermions.

As usual the equations satisfied by vector and axial-vector decouple from the other fields when one considers massless fermions. The kinetic equation in the presence of unique gauge field is well known \cite{hpy1,hsjlz,hpy2}. We generalize it and obtain the semiclassical chiral kinetic equation where the electromagnetic fields and the vorticity are treated on the same footing. 
By integrating the semiclassical relativistic kinetic equation over $p_0,$ we established a 3D CKT which does not depend on the spatial coordinates explicitly.  It is consistent with chiral anomaly and takes into account the non inertial properties of rotating reference frame. Moreover,  the chiral magnetic and vorticity effects are correctly generated. 

Transport equations of massive fermions are also studied. The semiclassical kinetic equations of the vector and axial-vector fields are derived  by extending the formalism given in \cite{hhy}. The related 3D kinetic transport equation is obtained by letting the spin 4-vector be  adequate to discuss the small mass limit. We showed that the Coriolis force and the dispersion relation are correctly generated.  We obtained the particle number current density in terms of the equilibrium distribution function and  calculated it  at   zero temperature.  The similarities and differences with the other approaches are discussed.  

For massive fermions we obtained the 3D transport equations by adopting the definition of spin vector as in  (\ref{amu})  and keeping only the linear terms in the electromagnetic fields and vorticity. For having a  better understanding of the mass corrections to the chiral effects, one needs to find a solution of the spin vector depending on the electromagnetic fields and vorticity. 

Although we considered the collisionless case, kinetic transport equations are mainly needed  in the presence of collisions. They  can be introduced in the current formalism by generalizing the methods given in \cite{hpy1} and \cite{yhh-col}. Obviously collisions can also be studied within  the 3D kinetic transport theories.

Zilch vortical effect \cite{maximZilch} has been  recently  studied  within the  kinetic theories  in \cite{hmss,hhyy}. In \cite{hmss} it was shown that the  zilch current in a rotating system can equivalently be derived  in chiral kinetic theory by employing  chiral current.
It would be interesting to inspect if this construction of zilch vortical effect can be studied by means of our approach for example by modifying   the chiral current.  This may provide an alternative way of studying the rotation of reference frame for photonic media.

\acknowledgments{ E.K. is partially supported by the Bogazici University Research Fund under Grant No. 20B03SUP3.}

\appendix

\section{ Integration of (\ref{cke12}) over $\mathbf{ p_0}$}
\label{appe}

Expressing the kinematic vorticity tensor $\Omega_{\mu \nu}$ in terms of the the fluid vorticity as in (\ref{Ommunu}) one can show that
$
S^{\mu \nu} \Omega_{\mu \nu}=p\cdot \omega/p_0
$
and 
$
\tilde{\Omega}^{\ssmn}=\omega^\mu u^\nu -\omega^\nu u^\mu.
$
Thus  (\ref{cke12}) can be rewritten as
\begin{eqnarray}
&&\delta\left( p^2 - \hbar \chi Q\frac{ p \cdot B}{u\cdot p}\right)   
\{ p \cdot \tilde{\nabla}  \nonumber\\
&& + \frac{\hbar \chi Q }{u \cdot p} S^{\mu \nu}E_{\mu }  \tilde{\nabla}_\nu  
- \frac{\hbar \chi}{u \cdot p}(p\cdot\omega\ u^\mu -u \cdot p\ \omega^\mu ) \tilde{\nabla}_\mu 
\label{cke3}\\
&& 
+(k-1)\frac{\hbar \chi  }{2(u \cdot p)^2}\left[p\cdot \omega \ p_{\perp}^{\mu} +(u \cdot p)^2 \omega^\mu \right]
\tilde{\nabla}_\mu  
\} f_\chi =0, \nonumber
\end{eqnarray}
where 
$
p_{\perp}^{\mu}=p^\mu -u \cdot p\, u^\mu.
$
It is  perpendicular to the fluid velocity:  $p_{\perp}\cdot u=0.$ 
 Integration of (\ref{cke3}) over $p_0$ in the frame  $u^\mu=(1,\bm 0)$ and  $\omega^\mu =(0, \bm \omega),$ leads to 
\begin{equation}
\label{3DT1}
\begin{aligned}
&\Big( \sqrt{\eta}_{\, s}^{\, \sschi } \frac{\partial }{\partial t} + (\sqrt{\eta} \dot{{\mathbf{ x}}})^\sschi_s  \cdot \frac{\partial }{\partial \mathbf{x}} 
\Big) f_{\sschi}^{s} (t,\mathbf x,\mathbf p) & \\
&+ (\sqrt{\eta}  \dot{\mathbf p})^\sschi_s \cdot \Big[ \frac{\partial }{\partial \mathbf{p}} f_{\sschi}^{s} (t,\mathbf x,\mathbf p, p_0)\Big]_{p_0=s{\cal E}_s^\sschi}
+I_0 \Big[ \frac{\partial }{\partial p_0} f_{\sschi}^{s} (t,\mathbf x,\mathbf p, p_0)\Big]_{p_0=s{\cal E}_s^\sschi}=0, & 
\end{aligned}
\end{equation}
where 
the phase space  measure $\sqrt{\eta}_{\, s}^{\, \sschi }$ and the velocities $(\sqrt{\eta} \dot{{\mathbf{ x}}})^\sschi_s,(\sqrt{\eta} \dot{{\mathbf{ p}}})^\sschi_s $ are given in (\ref{3e1})-(\ref{3e3}). 
The 3D distribution function is defined by
\begin{equation}
f_{\sschi}^{s} (t,\mathbf x,\mathbf p) \equiv f_{\sschi}^{s} (t,\mathbf x,\mathbf p, p_0)|_{p_0=s{\cal E}_s^\sschi}. \nonumber
\end{equation}
Coefficient of  the last term in (\ref{3DT1}) is 
\begin{eqnarray}
I_0 = \hbar\chi Q\frac{\bm \omega \cdot \bm E}{2|\bm p|} +\hbar\chi Q\frac{\bm p \cdot \bm\omega\, \bm p \cdot \bm E}{2|\bm p|^3} 
+ sQ\frac{\bm p\cdot \bm E}{|\bm p |}\left(1+\hbar \chi Q\frac{\bm p \cdot \bm B}{|\bm p|^3}{|\bm p|^2}\right) .\label{delp1}
\end{eqnarray}
A well defined 3D equation can be established  only  if  some terms in   (\ref{3DT1}) can be combined to yield
\begin{eqnarray}
\begin{aligned}
&(\sqrt{\eta}  \dot{\mathbf p})^\sschi_s \cdot \Big[ \frac{\partial }{\partial \mathbf{p}} f_{\sschi}^{s} (t,\mathbf x,\mathbf p, p_0)\Big]_{p_0=s{\cal E}_s^\sschi}
+(\sqrt{\eta} \dot{\bm p})^\sschi_s \cdot \frac{\partial {\cal E}^\sschi_s}{\partial \bm p}\Big[ \frac{\partial }{\partial p_0} f_{\sschi}^{s}(t,\mathbf x,\mathbf p, p_0) \Big]_{p_0=s{\cal E}_s^\sschi} \nonumber &\\&= (\sqrt{\eta}  \dot{\mathbf p})^\sschi_s \cdot \frac{\partial }{\partial \mathbf{p}} f_{\sschi}^{s} (t,\mathbf x,\mathbf p).&
\end{aligned}
\end{eqnarray}
For this purpose let us express the first term of (\ref{delp1}) in a  different form. First observe that \cite{dk} 
\begin{equation}
\int d^4p \ \delta (p^2)\Big\{ E \cdot \omega \left(\frac{f^{0s}_{\sschi}}{p_0}-\frac{1}{2}\frac{\partial f^{0s}_{\sschi} }{\partial p_0} \right)
+\frac{\omega\cdot  p  E \cdot  p }{p_0^2} \left(\frac{2f^{0s}_{\sschi}}{p_0}-\frac{1}{2}\frac{\partial f^{0s}_{\sschi} }{\partial p_0} \right) \Big\}=0, \label{conf0} 
\end{equation}
where
 \begin{equation}
f^{0s}_{\sschi} = \frac{2}{(2 \pi \hbar)^3}    \frac{1}{ e^{s( u \cdot p - \mu_\sschi) / T } +1  } \cdot
\label{f_fd1}
\end{equation}
Integration of (\ref{conf0}) over $p_0$  yields
\begin{equation}
\frac{\bm \omega \cdot \bm E}{2|\bm p|} \frac{\partial f^{0s}_{\sschi} }{\partial p_0}\Big|_{p_0=s|\bm p|} =
\frac{\bm p \cdot \bm\omega\, \bm p \cdot \bm E}{2|\bm p|^3} \frac{\partial f^{0s}_{\sschi} }{\partial p_0}\Big|_{p_0=s|\bm p|} +s\Big(\frac{\bm \omega \cdot \bm E}{|\bm p|^2} -2 \frac{\bm p \cdot \bm\omega\, \bm p \cdot \bm E}{|\bm p|^4}\Big) f^{0s}_{\sschi}(t,\mathbf x,\mathbf p),
\label{ome}
\end{equation}
with the shorthand notation
$$
\frac{\partial f^{s}_{\sschi} }{\partial p_0}\equiv \frac{\partial f^{s}_{\sschi} (t,\mathbf x,\mathbf p, p_0)}{\partial p_0}.
$$ 
Then by substituting the first term of (\ref{delp1}) with  (\ref{ome}) we  can write the last term of   (\ref{3DT1})  as
\begin{eqnarray}
I_0\frac{\partial f^{s}_{\sschi} }{\partial p_0}\Big|_{p_0=s{\cal E}_s^\sschi} &=&
\hbar\chi Q\frac{\bm p \cdot \bm\omega\, \bm p \cdot \bm E}{|\bm p|^3} \frac{\partial f^{s}_{\sschi} }{\partial p_0}\Big|_{p_0=s|\bm p|} +\hbar s\chi Q\Big(\frac{\bm \omega \cdot \bm E}{|\bm p|^2} -2 \frac{\bm p \cdot \bm\omega\, \bm p \cdot \bm E}{|\bm p|^4}\Big) f^{s}_{\sschi}(t,\mathbf x,\mathbf p)\nonumber\\
&&+Q \frac{\bm p\cdot \bm E}{|\bm p |}\left(1+\hbar \chi Q\frac{\bm p \cdot \bm B}{|\bm p|^3}{|\bm p|^2}\right)  \frac{\partial f^{s}_{\sschi} }{\partial p_0}\Big|_{p_0=s{\cal E}_s^\sschi} \cdot
\label{io}
\end{eqnarray}
On the other hand from (\ref{canvel}) and (\ref{3e3}) we have 
\begin{eqnarray}
(\sqrt{\eta} \dot{\bm p})^\sschi_s\cdot  \frac{\partial {\cal E}^\sschi_s}{\partial \bm p} = \hbar Q\chi \frac{\bm p\cdot \bm \omega\, \bm E\cdot \bm p}{|\bm p|^3}+ \frac{Q\bm p\cdot \bm E}{|\bm p |}\left(1+\hbar \chi Q\frac{\bm p \cdot \bm B}{|\bm p|^3}{|\bm p|^2}\right) .
\label{pdv}
\end{eqnarray}
Therefore, by making use of (\ref{io}) and (\ref{pdv}), we conclude that
\begin{eqnarray}
(\sqrt{\eta}  \dot{\mathbf p})^\sschi_s \cdot \Big[ \frac{\partial }{\partial \mathbf{p}} f_{\sschi}^{s} (t,\mathbf x,\mathbf p, p_0)\Big]_{p_0=s{\cal E}_s^\sschi}
+I_0 \Big[ \frac{\partial }{\partial p_0} f_{\sschi}^{s} (t,\mathbf x,\mathbf p, p_0)\Big]_{p_0=s{\cal E}_s^\sschi}=\nonumber \\
\Big[(\sqrt{\eta}  \dot{\mathbf p})^\sschi_s \cdot \frac{\partial }{\partial \mathbf{p}} 
+\hbar s\chi Q\Big(\frac{\bm \omega \cdot \bm E}{|\bm p|^2} -2 \frac{\bm p \cdot \bm\omega\, \bm p \cdot \bm E}{|\bm p|^4}\Big)\Big] f_{\sschi}^{s} (t,\mathbf x,\mathbf p) \label{pI0}
\end{eqnarray}
Plugging (\ref{pI0}) into (\ref{3DT1})  leads to (\ref{CKT3}).

\bibliography{QKE-fluid-JHEP-rev.bib}
\end{document}